\documentclass[12pt]{article}

\usepackage{latexsym}
\usepackage{amsthm}
\usepackage{verbatim}
\usepackage[dutch,english]{babel}
\usepackage{amsfonts}
\usepackage{mathtools}
\usepackage{bbm}
\usepackage{lscape}
\usepackage{amsmath,amsfonts,amssymb}
\usepackage{amsmath,amsbsy,amssymb,latexsym, graphics, graphicx, tabularx}
\usepackage{multirow,multicol,subfigure,booktabs}
\usepackage{algorithm}
\usepackage{lscape}
\usepackage{color}
\usepackage{url}
\usepackage{rotating}
\usepackage{graphicx}
\usepackage{subfig}
\usepackage{natbib}
\usepackage{longtable}

\newtheorem{Lemma}{Lemma}[section] 
\newtheorem{Theorem}{Theorem}[section] 

\makeatletter
\renewcommand\theequation{\thesection.\arabic{equation}}
\@addtoreset{equation}{section}
\makeatother

\usepackage[a4paper, total={6in, 8in}]{geometry}
\renewcommand{\baselinestretch}{1.2}

\newcommand{\C}{{\cal C}}
\newcommand{\D}{{\cal D}}
\begin{document}

\newcommand{\tcr}{\color{black}}
\newcommand{\tcb}{\color{black}}

\title{\bf Dependent censoring based on copulas}

\author{
{\large Claudia C\textsc{{zado}}
\footnote{Technical University of Munich, E-mail address: \texttt{cczado@ma.tum.de}. Claudia Czado acknowledges the support of the German Science Foundation
(DFG grant CZ 86/6-1)} } 
\\
\and
\addtocounter{footnote}{2} 
{\large Ingrid V\textsc{{an} K{eilegom}}
\footnote{ORSTAT, KU Leuven, E-mail address: \texttt{ingrid.vankeilegom@kuleuven.be}. Ingrid Van Keilegom acknowledges the support of the European Research Council (2016-2021, Horizon 2020 / ERC grant agreement No.\ 694409).} } \\
}

\date{\today}

\maketitle

\begin{abstract}
Consider a survival time $T$ that is subject to random right censoring, and suppose that $T$ is stochastically dependent on the censoring time $C$.  We are interested in the marginal distribution of $T$.   This situation is often encountered in practice.  Consider for instance the case where $T$ is the time to death of a patient suffering from a certain disease.  Then, the censoring time $C$ is for instance the time until the person leaves the study or the time until he/she dies from another disease.  If the reason for leaving the study is related to the health condition of the patient or if he/she dies from a disease that has similar risk factors as the disease of interest, then $T$ and $C$ are likely dependent.   In this paper we propose a new model that takes this dependence into account.  The model is based on a parametric copula for the relationship between $T$ and $C$, and on parametric marginal distributions for $T$ and $C$.   Unlike most other papers in the literature, we do not assume that the parameter defining the copula function is known.   We give sufficient conditions on these parametric copula and marginals under which the bivariate distribution of $(T,C)$ is identified.   These sufficient conditions are then checked for a wide range of common copulas and marginal distributions.   We also study the estimation of the model, and carry out extensive simulations and the analysis of data on pancreas cancer to illustrate the proposed model and estimation procedure.  
\end{abstract}


\def\baselinestretch{1.3}
\newpage
\normalsize
\setcounter{footnote}{0}
\setcounter{equation}{0}

\section{Introduction}  \label{sect1}

A very common situation in survival analysis is that duration times are right censored.  This can happen for several reasons.  In medical studies it often happens that patients who are followed over time until they die of a certain disease are still alive at the end of the study, they leave the study before the end for various reasons, or they die from another cause.   In most papers in the literature it is assumed that the survival time $T$ is independent of the censoring time $C$, where independent censoring should here be interpreted as stochastic independence.   However, there are many instances in which this independence assumption is violated.   Consider for instance the case where the patient leaves the study for reasons related to his/her health, or where he/she dies of another related disease.  In these cases it is important to take the dependence between $T$ and $C$ into account in the model.  However, the seminal paper by \citet{Tsiatis75} shows that the bivariate distribution of $T$ and $C$ is not identifiable in a completely nonparametric setting.   Some authors have therefore proposed parametric or semiparametric models for this bivariate distribution that are identifiable.  A popular model in that context is the copula model, that allows to model the marginal laws of $T$ and $C$ separately from the relation between $T$ and $C$.   The first paper in that context was \citet{Zheng95}.   They worked with a completely known copula, and proposed a nonparametric estimator of the marginal distribution of $T$ and $C$, called a copula-graphic estimator, that generalizes the \citet{Kaplan58} estimator to the case of dependent censoring.  Later on, \citet{Rivest01} focused on the case of Archimedean copulas and obtained a closed-form expression for the copula-graphic estimator.   However, a major drawback of this estimator is that it relies on a completely known copula, for instance a Frank copula with {\it known} association parameter.  In practice the association parameter is often not known, and can have a major influence on the resulting estimator of the marginal distributions.   Other contributions in the same vein (with known copula) are given in \citet{Braekers05}, \citet{Huang08}, \citet{Chen10}, \citet{deUA16}, \citet{Sujica18} and \citet{Emura18}, among others, whereas other approaches not based on copulas have been studied by \citet{Nadas71}, \citet{Basu78}, \citet{Basu88}, \citet{Emoto90}, \citet{Scharfstein02}, \citet{Jackson14}, \citet{Collett15}, \citet{Hsu15}, and \citet{Deresa20a, Deresa20b, Deresa20c}, among others.  \tcr The latter papers only consider rather specific (semi-)parametric model assumptions, that can not be easily extended to other contexts. \tcb

In this paper we will show that the assumption of a completely known copula is not a necessary condition to identify the joint distribution of $T$ and $C$.  We will show that if the marginal distributions of $T$ and $C$ and the copula function are all modelled parametrically, then under certain conditions the joint model is identifiable.  In particular, the association parameter of the copula function is identifiable, which is an important step forward in the use of copulas in survival analysis.   This might seem surprising, since under right censoring one either observes $T$ or $C$, but in general not both.  We will show that the identification of the relation between $T$ and $C$ does not follow using standard lines of reasoning commonly used in parametric models, but instead it is a delicate exercise that makes use of the available information in an optimal way.  The price to pay for the identifiability of the association parameter is that the marginals are no longer fully nonparametric.   We believe however that this is an acceptable price, given that one often has no idea how to choose the association parameter in practice.  We will develop sufficient conditions on the families of marginal distributions and on the family of copula functions under which the joint model is identifiable.   These sufficient conditions are satisfied for a wide range of common parametric marginal distributions and copula functions, making the model very useful in practice.

The paper is organized as follows.  In the next section we introduce the copula model, and some notations, definitions and formulas that are needed later in the paper.   Section \ref{sect3} develops sufficient conditions under which the model is identifiable.  The estimation of the model is studied in Section \ref{sect4}.  Sections \ref{sect5} and \ref{sect6} consider the results of extensive simulations and of the analysis of data on pancreas cancer, respectively.  Some ideas for further research are discussed in Section \ref{sect7}, while the proofs are collected in the Appendix.

\section{The model, and some notations and definitions}  \label{sect2}

Let $T$ be a survival time, and $C$ be a censoring time.  Due to random right censoring we observe $Y=\min(T,C)$ and $\Delta=I(T \le C)$.  We allow for stochastic dependence between $T$ and $C$ and will model this dependence with a copula.  Throughout the paper we assume that $T$ and $C$ are non-negative and that the marginal distributions $F_T$ and $F_C$ of $T$ and $C$ are continuous and belong to parametric families:
\begin{align} \label{model1}
F_T \in \{F_{T,\theta_T} : \theta_T \in \Theta_T\}, \quad \mbox{and} \quad F_C \in \{F_{C,\theta_C} : \theta_C \in \Theta_C\}, 
\end{align}
for certain parameter spaces $\Theta_T$ and $\Theta_C$.  We denote their densities by $f_T$ and $f_C$ (or $f_{T,\theta_T}$ and $f_{C,\theta_C}$ in the parametric families).  We model the bivariate distribution $F_{T,C}$ of $(T,C)$ by a copula based model.  A copula is a bivariate distribution function $\C : [0,1] \times [0,1] \rightarrow [0,1]$ with uniform margins.  Thanks to \citet{Sklar59} and the continuity of the marginal distributions $F_T$ and $F_C$, we know that there is a unique copula $\C$ for which 
\begin{align} \label{model1bis}
F_{T,C}(t,c) = \C(F_T(t),F_C(c))  
\end{align}
holds for any $t, c \ge 0$.  We model the copula parametrically:
\begin{align} \label{model2}
\C \in \{\C_\theta : \theta \in \Theta\}, 
\end{align}
for some parameter space $\Theta$.  Note that instead of (\ref{model1bis})-(\ref{model2}) we could also assume that $P(T>t,C>c) = \tilde \C(1-F_T(t),1-F_C(c))$ with $\tilde \C$ belonging to a certain parametric class of survival copulas.   There is no substantial difference between the two ways of modelling the copula.   We will focus on the usual copulas, but survival copulas can be used as  an alternative.    

For our approach we need to express conditional distribution functions in terms of their associated copulas and define
$$ h_{C|T,\theta}(v|u) = \frac{\partial}{\partial u} \C_\theta(u,v) \quad \mbox{and} \quad h_{T|C,\theta}(u|v) = \frac{\partial}{\partial v} \C_\theta(u,v). $$
Further we let $F_{Y,\Delta}(y,\delta) = P(Y \le y,\Delta=\delta)$ and $f_{Y,\Delta}(y,\delta) = (d/dy)F_{Y,\Delta}(y,\delta)$ for $\delta=0,1$.  Finally, let $F_Y(y) = \sum_{\delta=0}^1 F_{Y,\Delta}(y,\delta)$ and $f_Y(y) = \sum_{\delta=0}^1 f_{Y,\Delta}(y,\delta)$ be the distribution and density of the observable random variable $Y$, respectively.  Note that we can express
\begin{align}
& F_{T|C}(t|c) = h_{T|C}(F_T(t)|F_C(c)), \quad \quad \quad F_{C|T}(c|t) = h_{C|T}(F_C(c)|F_T(t)), \nonumber \\
& F_Y(y) = F_T(y) + F_C(y) - \C(F_T(y),F_C(y)), \nonumber\\
& f_{Y,\Delta}(y,1) = f_T(y) [1-h_{C|T}(F_C(y)|F_T(y))], \label{resultsYdelta} \\
& f_{Y,\Delta}(y,0) = f_C(y) [1-h_{T|C}(F_T(y)|F_C(y))]. \nonumber 
\end{align}
For a derivation of the results given in \eqref{resultsYdelta} we refer to the Appendix.
For all functions introduced above, copula parameters $\theta$ and marginal parameters $\theta_T$ and $\theta_C$ can be added to indicate the parametric versions of these functions, for instance with $\boldsymbol{\alpha} = (\theta,\theta_T,\theta_C)^\top$, we denote $f_{Y,\Delta,\boldsymbol{\alpha}}(y,1) = f_{T,\theta_T}(y) [1-h_{C|T,\theta}(F_{C,\theta_C}(y)|F_{T,\theta_T}(y))]$.


\section{Identifiability} \label{sect3}

The study of the identifiability of model (\ref{model1})-(\ref{model2}) is the backbone of this paper, and we investigate it in detail in this section.  All proofs are given in the Appendix. 

We start with a general result that gives sufficient conditions under which the model is identifiable.  With identifiability we mean that the parameters $(\theta,\theta_T,\theta_C) \in \Theta \times \Theta_T \times \Theta_C$ determine in a unique way the density of the observable random variables $(Y,\Delta)$, i.e.\ if $f_{Y,\Delta,\boldsymbol{\alpha}_1} \equiv f_{Y,\Delta,\boldsymbol{\alpha}_2}$, then $\boldsymbol{\alpha}_1=\boldsymbol{\alpha}_2$, where $\boldsymbol{\alpha}_j=(\theta_j,\theta_{Tj},\theta_{Cj})^\top$, $j=1,2$.  

\begin{Theorem} \label{ident}
Suppose that 
\begin{itemize}
\item[(C1)] For $a=0$ and $a=\infty$, and for $\theta_{T1},\theta_{T2} \in \Theta_T$ and $\theta_{C1},\theta_{C2} \in \Theta_C$, we have that
$$ \lim_{t \rightarrow a} \frac{f_{T,\theta_{T1}}(t)}{f_{T,\theta_{T2}}(t)}=1 \Longleftrightarrow \theta_{T1} = \theta_{T2}, $$ 
and
$$ \lim_{t \rightarrow a} \frac{f_{C,\theta_{C1}}(t)}{f_{C,\theta_{C2}}(t)}=1 \Longleftrightarrow \theta_{C1} = \theta_{C2}. $$ 
\item[(C2)]  
The parameter space $\Theta \times \Theta_T \times \Theta_C$ is such that 
$$ (C2a): \quad \min \Big\{{\max}^*\lim_{t \rightarrow 0} h_{T|C,\theta}(u_t|v_t), {\max}^*\lim_{t \rightarrow \infty} h_{T|C,\theta}(u_t|v_t)\Big\} = 0, $$
and
$$ (C2b): \quad \min \Big\{{\max}^*\lim_{t \rightarrow 0} h_{C|T,\theta}(v_t|u_t), {\max}^*\lim_{t \rightarrow \infty} h_{C|T,\theta}(v_t|u_t)\Big\} = 0, $$
where $u_t = F_{T,\theta_T}(t), v_t = F_{C,\theta_C}(t)$ and ${\max}^*$ is the maximum over all $(\theta,\theta_T,\theta_C) \in \Theta \times \Theta_T \times \Theta_C$. 
\end{itemize}
Then, model (\ref{model1})-(\ref{model2}) is identified.
\end{Theorem}

Condition (C1) is valid for a wide range of parametric families for the marginal densities $f_T$ and $f_C$, as is shown in the next theorem.  For other families not mentioned below, condition (C1) can be easily checked as well, but we restrict attention to the most important parametric families in survival analysis.

\begin{Theorem} \label{condC1}
Condition (C1) is satisfied for the families of log-normal, log-Student-t, Weibull, and log-logistic densities.   
\end{Theorem}

We will now study two important classes of copulas in more detail, namely \linebreak Archimedean copulas and Gaussian copulas.  We will verify in which cases  condition (C2) is valid for these classes.  Archimedean copulas can be written as 
\begin{align} \label{cop}
& \C(u,v) = \psi^{[-1]}(\psi(u) + \psi(v)), 
\end{align}
where $\psi$ is a generator, i.e.\  $\psi : [0,1] \rightarrow [0,\infty)$ is a continuous, strictly decreasing and convex function such that $\psi(1) = 0$.  Here, $\psi^{[-1]}$ is the pseudo-inverse of $\psi$, i.e.\ $\psi^{[-1]}(t) = \psi^{-1}(t)$ if $0 \le t \le \psi(0)$ and $\psi^{[-1]}(t) = 0$ if $t \ge \psi(0)$.  Important families of Archimedean copulas are the Frank family, corresponding to 
\begin{align*}
& \psi_\theta(u) = - \log \Big(\frac{e^{-\theta u}-1}{e^{-\theta}-1}\Big),
\end{align*}
with $\theta \in \mathcal{R} \backslash \{0\}$, the Clayton family for which $\psi_\theta(u) = \theta^{-1}(u^{-\theta}-1)$, with $\theta \in [-1,\infty) \backslash \{0\}$, and the Gumbel family, defined by $\psi_\theta(u) = (-\log(u))^\theta$, with $\theta \in [1,\infty)$.

Differentiation of (\ref{cop}) gives 
$$ h_{T|C}(u|v) = \frac{\psi'(v)}{\psi'\big(\psi^{-1}(\psi(u)+\psi(v))\big)} $$
and
$$ h_{C|T}(v|u) = \frac{\psi'(u)}{\psi'\big(\psi^{-1}(\psi(u)+\psi(v))\big)} $$
for $0 < u,v < 1$, provided the derivatives and inverses in this formula exist.

\tcr The following lemma helps in assessing whether condition (C2a) is verified or not.  A similar result exists for condition (C2b). 

\begin{Lemma} \label{lem1}
Suppose the generator $\psi$ is differentiable on $(0,1)$.
If $\lim_{v \rightarrow 1} \psi'(v)  \in (-\infty,0)$, then $\lim_{t \rightarrow \infty} h_{T|C,\theta}(F_{T,\theta_T}(t)|F_{C,\theta_C}(t)) = 1$.
\end{Lemma}
\tcb

Lemma \ref{lem1} can be used to assess whether (C2) holds true. Below we do this for some of the well known parametric Archimedean copula families.  We will also consider the Gaussian copula (which is not an Archimedean copula) defined by 
$$ \C_\theta(u,v) = \Phi_\theta(\Phi^{-1}(u),\Phi^{-1}(v)), $$  
where $\Phi$ is the cumulative distribution of a standard normal random variable, and $\Phi_\theta$ is the cumulative distribution of a bivariate standard normal random vector with correlation $\theta$.  

\begin{Theorem}{\bf \tcr[Frank, Gumbel and Gaussian copula]\tcb} \label{spec}
Condition (C2) is satisfied for 
\begin{itemize}
\item[(i)] the Frank copula, independently of the marginal distributions and the size of the parameter space $\Theta \times \Theta_T \times \Theta_C$.
\item[(ii)] the Gumbel copula if $\lim_{t \rightarrow 0} \log F_{T,\theta_T}(t)/\log F_{C,\theta_C}(t) \in (0,\infty)$ for all $(\theta_T,\theta_C) \in \Theta_T \times \Theta_C$.
\item[(iii)] for the Gaussian copula if 
$$ \min\Big\{{\max}^*\lim_{t \rightarrow 0} A_{\theta,F_{T,\theta_T},F_{C,\theta_C}}(t),{\max}^*\lim_{t \rightarrow \infty} A_{\theta,F_{T,\theta_T},F_{C,\theta_C}}(t)\Big\} = -\infty $$ 
and 
$$ \min\Big\{{\max}^*\lim_{t \rightarrow 0} A_{\theta,F_{C,\theta_C},F_{T,\theta_T}}(t),{\max}^*\lim_{t \rightarrow \infty} A_{\theta,F_{C,\theta_C},F_{T,\theta_T}}(t)\Big\} = -\infty, $$ 
where $A_{\theta,F_1,F_2}(t) = \Phi^{-1}(F_1(t)) - \theta \Phi^{-1}(F_2(t))$, and ${\max}^*$ is the maximum over all $(\theta,\theta_T,\theta_C) \in \Theta \times \Theta_T \times \Theta_C$. 
\end{itemize}
\end{Theorem}

Note that it can be easily seen that $\lim_{t \rightarrow 0} \log F_{T,\theta_T}(t)/\log F_{C,\theta_C}(t) = \sigma_C^2/\sigma_T^2$ if $\log T \sim N(\mu_T,\sigma_T^2)$ and $\log C \sim N(\mu_C,\sigma_C^2)$, and that this limit equals $\rho_T/\rho_C$ if $T \sim$ Weibull$(\lambda_T,\rho_T)$ and $C \sim$ Weibull$(\lambda_C,\rho_C)$ (we refer to (\ref{Weib}) for the definition of the Weibull parameters).  Hence, the Gumbel copula can be used for these two \tcr marginal specifications. \tcb

Regarding the Gaussian copula, note that for any distribution $F$, the function $\Phi^{-1}(F(t))$ tends to $-\infty$ and $+\infty$ when $t$ tends to zero and infinity, respectively.  Hence, the limit of $A_{\theta,F_1,F_2}(t)$ will be determined by which of the functions $F_1$ and $F_2$ dominates in the limit.  As an example, consider the case where $\log T \sim N(\mu_T,\sigma_T^2)$ and $\log C \sim N(\mu_C,\sigma_C^2)$.  Then, $\Phi^{-1}(F_T(t)) = (\log(t)-\mu_T)/\sigma_T$, and hence 
$$ A_{\theta,F_T,F_C}(t) = \big(\frac{1}{\sigma_T}-\frac{\theta}{\sigma_C}\big) \log(t) - \big(\frac{\mu_T}{\sigma_T}-\frac{\theta \mu_C}{\sigma_C}\big), $$
and this tends to $-\infty$ either when $t$ tends to zero or to infinity depending on the sign of $\sigma_C-\theta \sigma_T$.   \tcb This shows that for log-normal margins and for a Gaussian copula, condition (C2) is satisfied locally around the true parameter values.  \tcb See also \cite{Nadas71}, \citet{Basu78} and \citet{Deresa20a}, who considered the bivariate normal case.   \tcb For other margins the condition is harder to verify, since the normal quantile function $\Phi^{-1}$ is difficult to handle.  Numerical calculation of the function $A_{\theta,F_T,F_C}(t)$ when $T \sim$ Weibull$(\lambda_T,\rho_T)$ and $C \sim$ Weibull$(\lambda_C,\rho_C)$ \tcr supports \tcb however that the limit of this function equals $-\infty$ if $\rho_T < \rho_C$ and $t$ tends to $\infty$, or if $\rho_T > \rho_C$ and $t$ tends to 0.  \tcb

Condition (C2) is however a sufficient condition for identification, but not per se a necessary condition.  It only uses the limits when $t$ goes to zero or infinity, whereas the identification proof can also use all other time points in between.   We will illustrate this for the Clayton copula with $\theta >0$ (in which case the Clayton copula is strict, \tcr i.e.\ $\psi(0)=\infty$). \tcb  
One can easily show that for the Clayton copula we have that (see also \cite{Aas09})
\begin{align*}
& h_{T|C,\theta}(F_T(t)|F_C(t)) = \Big[1+\Big(\frac{F_C(t)}{F_T(t)}\Big)^\theta - F_C(t)^\theta\Big]^{-\frac{\theta+1}{\theta}}, 
\end{align*}
and 
\begin{align} \label{hCT}
& h_{C|T,\theta}(F_C(t)|F_T(t)) = \Big[1+\Big(\frac{F_T(t)}{F_C(t)}\Big)^\theta - F_T(t)^\theta\Big]^{-\frac{\theta+1}{\theta}}. 
\end{align}
Hence, it is not possible that both functions tend to zero when $t$ tends to zero.  Since they both converge to 1 when $t$ tends to infinity, we see that condition (C2) is not satisfied.   
However, from the fact that $f_{Y,\Delta}(y,0) = f_C(y) \{1-h_{T|C}(F_T(y)|F_C(y))\}$ for all $y$ and similarly for $f_{Y,\Delta}(y,1) = f_T(y) \{1-h_{C|T}(F_C(y)|F_T(y))\}$, we can identify the parameters $\theta, \theta_T$ and $\theta_C$, as the following theorem shows.

\begin{Theorem}{\bf \tcr [Clayton copula] \tcb} \label{Clay}
Suppose that (C1) holds, that $\Theta_T \times \Theta_C$ is such that $\lim_{t \rightarrow 0} F_{T,\theta_T}(t)/F_{C,\theta_C}(t)$ is either 0 or $+\infty$ for all $\theta_T \in \Theta_T$ and $\theta_C \in \Theta_C$, and that the copula $\C_\theta$ is a Clayton copula with $\theta>0$.  Then, model (\ref{model1})-(\ref{model2}) is identified.
\end{Theorem}

Note that the condition that $\lim_{t \rightarrow 0} F_{T,\theta_T}(t)/F_{C,\theta_C}(t) = 0$ or $+\infty$ is satisfied for many parametric families.   Consider e.g.\ the log-normal family for $T$ and $C$, i.e.\ $\log T \sim N(\mu_T,\sigma_T^2)$ and $\log C \sim N(\mu_C,\sigma_C^2)$.  Then,
\begin{align*}
& \lim_{t \rightarrow 0} \frac{F_T(t)}{F_C(t)} = \lim_{t \rightarrow 0} \frac{f_T(t)}{f_C(t)} \\
& = \frac{\sigma_C}{\sigma_T} \lim_{t \rightarrow 0} \exp\Big\{-\frac12 \frac{(\log t-\mu_T)^2}{\sigma_T^2} + \frac12 \frac{(\log t-\mu_C)^2}{\sigma_C^2} \Big\} \\
& = \frac{\sigma_C}{\sigma_T} \lim_{t \rightarrow 0} \exp\Big\{-\frac12 \Big(\frac{1}{\sigma_T^2}- \frac{1}{\sigma_C^2}\Big) (\log t)^2 + \Big(\frac{\mu_T}{\sigma_T^2}-\frac{\mu_C}{\sigma_C^2}\Big) \log t - \frac12 \Big(\frac{\mu_T^2}{\sigma_T^2}-\frac{\mu_C^2}{\sigma_C^2}\Big) \Big\}.
\end{align*}
This converges to 0 if $\sigma_T < \sigma_C$ or if $\sigma_T=\sigma_C$ and $\mu_T>\mu_C$.  Similarly, the expression converges to infinity if $\sigma_T > \sigma_C$ or if $\sigma_T=\sigma_C$ and $\mu_T < \mu_C$.  \tcb  Hence, the model is identified locally around the true $(\mu_T,\sigma_T)$ and $(\mu_C,\sigma_C)$. \tcb   Note that if $\mu_T=\mu_C$ and $\sigma_T=\sigma_C$, it is easily seen that $h_{T|C}(F_T(t)|F_C(t)) = (2-F_C(t)^\theta)^{-(\theta+1)/\theta}$ for the Clayton copula.  Hence, $\lim_{t \rightarrow 0} h_{T|C}(F_T(t)|F_C(t)) = 2^{-(\theta+1)/\theta}$.  It follows that $\lim_{t \rightarrow 0} f_{C,\mu_{C1},\sigma_{C1}}(t) / f_{C,\mu_{C2},\sigma_{C2}}(t)$ $= (1-2^{-(\theta_2+1)/\theta_2}) / (1-2^{-(\theta_1+1)/\theta_1})$ for two sets of parameters $(\theta_1,\mu_{C1},\sigma_{C1})$ and $(\theta_2,\mu_{C2},\sigma_{C2})$.  Since the limit $\lim_{t \rightarrow 0} f_{C,\mu_{C1},\sigma_{C1}}(t) / f_{C,\mu_{C2},\sigma_{C2}}(t)$ can only be equal to 0, 1 or $\infty$ for the log-normal density, it follows that $\theta_1=\theta_2$ and hence $(\mu_{C1},\sigma_{C1}) = (\mu_{C2},\sigma_{C2})$.   This shows that the model is also identifiable when $\mu_T=\mu_C$ and $\sigma_T=\sigma_C$.  

\tcb Using similar (but easier) calculations we can show that if $T \sim$ Weibull$(\lambda_T,\rho_T)$ and $C \sim$ Weibull$(\lambda_C,\rho_C)$, then $\lim_{t \rightarrow 0} \frac{F_T(t)}{F_C(t)} = 0$ if $\rho_T > \rho_C$, it equals $\infty$ if $\rho_T < \rho_C$ and it equals $\lambda_T/\lambda_C$ if $\rho_T=\rho_C$ (we refer to (\ref{Weib}) for the definition of the Weibull parameters). \tcb

\section{Estimation}\label{sect4}

We now consider parameter estimation of the joint parametric model for the survival time $T$ and the censoring time $C$ specified in \eqref{model1}-\eqref{model2}. For this we assume that we have an i.i.d.\ sample $\D=\{(y_i,\delta_i), i=1,\ldots,n\}$ available. Then the joint log-
likelihood for the parameter vector $\boldsymbol{\alpha}=(\theta,\theta_T,\theta_C)^\top$ is given by
\begin{align}
   \ell(\boldsymbol{\alpha};\D) & = \sum_{i=1}^n \log \left(
   f_{Y,\Delta,\boldsymbol{\alpha}}(y_i,\delta_i) \right) \nonumber\\
   & = \sum_{i=1, \delta_i=1}^n
   \log \left(f_{T,\theta_T}(y_i) \{1-h_{C|T,\theta}(F_{C,\theta_C}(y_i)|F_{T,\theta_T}(y_i))\}
   \right) \nonumber\\
   & \quad + \sum_{i=1, \delta_i=0}^n
   \log \left(f_{C,\theta_C}(y_i) \{1-h_{T|C,\theta}(F_{T,\theta_T}(y_i)|F_{C,\theta_C}(y_i))\}   \right). \label{loglike}
\end{align}
We follow a maximum likelihood approach by maximizing the log-likelihood specified in \eqref{loglike}, i.e.\ we define parameter estimators by
$$ \boldsymbol{\hat\alpha} = (\hat\theta,\hat\theta_T,\hat\theta_C)^\top = \mbox{argmax}_{\boldsymbol{\alpha} \in A}  \ell(\boldsymbol{\alpha};\D), $$
where $A = \Theta \times \Theta_T  \times \Theta_C$.  
If we use for instance log-normal margins for $T$ and $C$ and single-parameter copula families we need to optimize over five parameters. In the simulation study and the data application we use unconstrained optimization. For standard one-parameter copula families such as the Frank, Clayton,  Gumbel and Gauss copula there exist a one-to-one relationship between Kendall's $\tau$ parameter and the copula parameter $\theta$. If there is positive dependence we use a logit transformation for $\tau$ and the Fisher's z transform of $\tau$ otherwise. So the Fisher's z transform of $\tau$ can be used in the Frank and Gauss copula case. This also facilitates a common interpretation of the dependence strength measured in terms of Kendall's $\tau$ over the different copula families.

In order to obtain the asymptotic normality of $(\hat\theta,\hat\theta_T,\hat\theta_C)$, we make use of \citet{White82}, who developed sufficient conditions under which a parameter estimator defined as the maximizer of a certain criterion function is asymptotically normal.   The results allow for misspecification of the parametric model.   Let $\boldsymbol{\alpha}^*=(\theta^*,\theta_T^*,\theta_C^*)^\top$ be the parameter vector that minimizes the Kullback-Leibler information criterion $E\big[\log f_{Y,\Delta}(Y,\Delta) - \log f_{Y,\Delta,\boldsymbol{\alpha}}(Y,\Delta) \big]$, and let $d = \mbox{dim}(\Theta) + \mbox{dim}(\Theta_T) + \mbox{dim}(\Theta_C)$.  Then, we have the following result. 

\begin{Theorem}  \label{asnorm}
(i) Under regularity conditions (A1) to (A3) in \citet{White82}, 
$$ (\hat\theta,\hat\theta_T,\hat\theta_C) \mathrel{\substack{P\\ \longrightarrow\\ n \rightarrow \infty}} (\theta^*,\theta_T^*,\theta_C^*). $$
(ii) Under regularity conditions (A1) to (A6) in \citet{White82}, 
$$ n^{1/2} \Big((\hat\theta,\hat\theta_T,\hat\theta_C) - (\theta^*,\theta_T^*,\theta_C^*) \Big) \mathrel{\substack{d\\ \longrightarrow\\ n \rightarrow \infty}} N(0,C), $$
where
$$ C = A(\boldsymbol{\alpha}^*)^{-1} B(\boldsymbol{\alpha}^*) A(\boldsymbol{\alpha}^*)^{-1}, $$
with  
\begin{align*}
A(\boldsymbol{\alpha}) &= \Big(E \Big\{\frac{\partial^2}{\partial \alpha_j \partial \alpha_k} \log f_{Y,\Delta,\boldsymbol{\alpha}}(Y,\Delta)\Big\}\Big)_{j,k=1}^d, \\
B(\boldsymbol{\alpha}) &= \Big(E \Big\{\frac{\partial}{\partial \alpha_j} \log f_{Y,\Delta,\boldsymbol{\alpha}}(Y,\Delta) \frac{\partial}{\partial \alpha_k} \log f_{Y,\Delta,\boldsymbol{\alpha}}(Y,\Delta)\Big\}\Big)_{j,k=1}^d. 
\end{align*}
\end{Theorem}

Note that, if the model is correctly specified, the matrix $C$ appearing in Theorem \ref{asnorm} equals $A(\boldsymbol{\alpha})^{-1}$, the inverse Fisher matrix.  

\section{Simulation study} \label{sect5}

We study the performance of the maximum likelihood estimators of $\theta$, $\theta_T$ and $\theta_C$ for four parametric copula families, namely the Frank, Clayton, Gumbel and Gauss copula, and for log-normal margins for $T$ and $C$, respectively. In particular the parameters of a log-normal random variable $X$ are $\mu$ and $\sigma$ and they correspond to the mean of $\log(X)$ and the standard error of $\log(X)$, respectively. Two different simulation scenarios are investigated with parameter specification given in Table \ref{tab:senarios}.      

\begin{table}[ht]
    \centering
    \begin{tabular}{rrrrrr}
    \hline
       Scenario  & $\mu_T$ & $\sigma_T$ & $\mu_C$ & $\sigma_C$ & $\tau$\\
       \hline
       1 & 2.2 & 1.0 & 2.0 & .25 & .2\\
       &&&&& .5\\
       &&&&& .8\\
       \hline
       2 &2.5 & 1.0 & 2.0 & .50 &.2\\
       &&&&& .5\\
       &&&&& .8\\
     \hline  
    \end{tabular}  
    \caption{Parameter specifications of the simulation scenarios with log-normal margins with parameters $\mu_T$ ($\mu_C$) and standard deviation $\sigma_T$ ($\sigma_C$) for $T$ and $C$, respectively, and dependence strength measured by Kendall's $\tau$. }
    \label{tab:senarios}
\end{table}

\begin{figure}[h!]
\centering
\includegraphics[scale=.52]{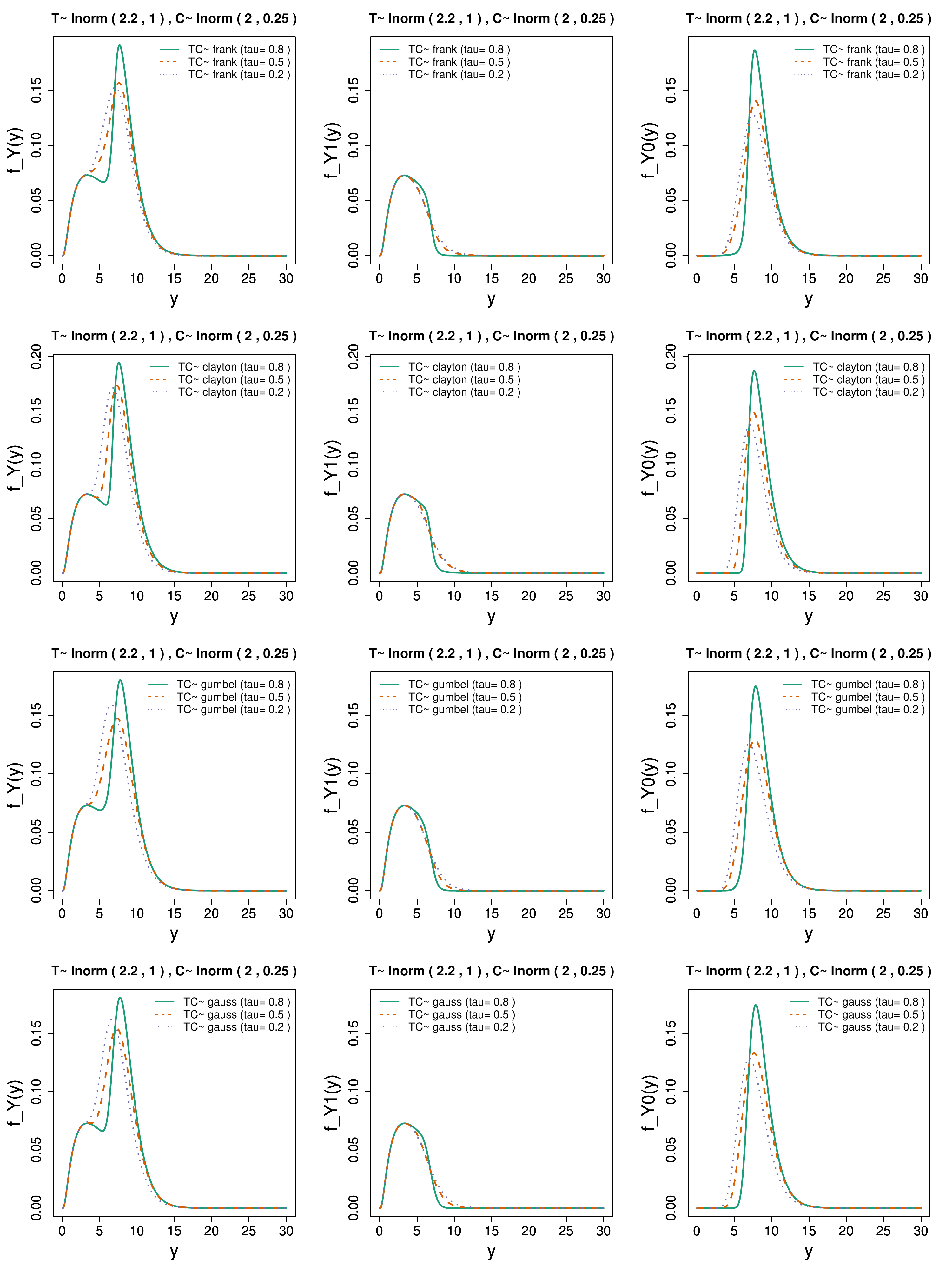}
\caption{Theoretical (sub)density $f_Y$ (left column), $f_{Y,\Delta}(\cdot,1)$ (middle column) and $f_{Y, \Delta}(\cdot,0)$ (right column) using a  Frank copula (top row), Clayton copula (second row), Gumbel copula (third row) and Gauss copula (bottom row) for Scenario 1.}
\label{fig:senario1}
\end{figure}

\begin{figure}[h!]
\centering
\includegraphics[scale=.52]{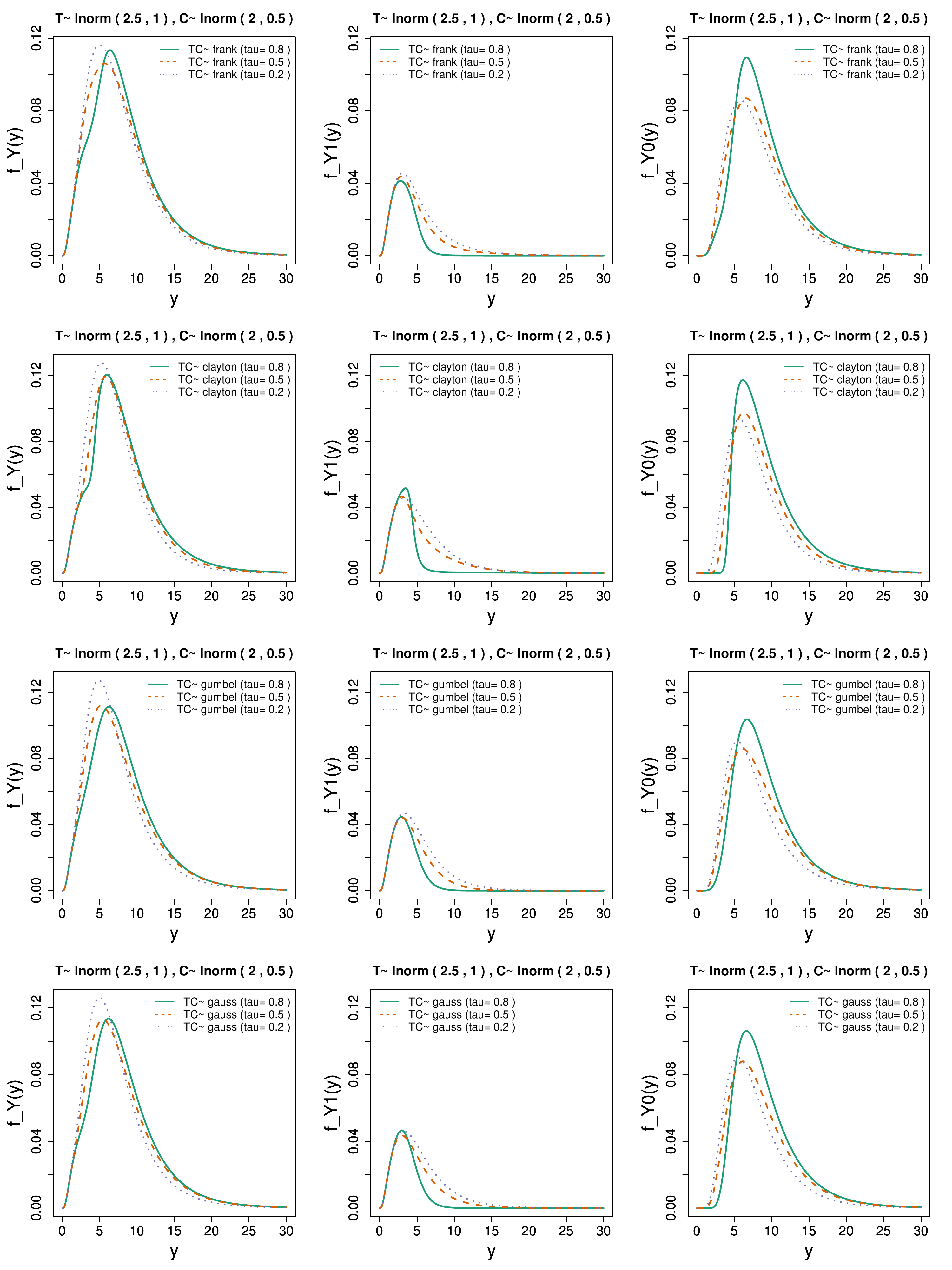}
\caption{Theoretical (sub)density $f_Y$ (left column), $f_{Y,\Delta}(\cdot,1)$ (middle column) and $f_{Y, \Delta}(\cdot,0)$ (right column) using a  Frank copula (top row), Clayton copula (second row), Gumbel copula (third row) and Gauss copula (bottom row) for Scenario 2.}
\label{fig:senario2}
\end{figure}

A visualization of the resulting theoretical (sub)densities $f_Y$, $f_{Y,\Delta}(\cdot,0)$ and $f_{Y,\Delta}(\cdot,1)$ from both scenarios based on the expressions given in \eqref{resultsYdelta} are given in Figures \ref{fig:senario1} and \ref{fig:senario2}, respectively.
Especially for Scenario 1 we see that the resulting marginal density of the observable random variable $Y$ is non standard. This is expected since it is the sum of two subdensities. Further we see that the strength of the dependence between $T$ and $C$ influences the skewness of $Y$.

The resulting uncensoring probabilities $P(\Delta=1)$ for all simulation scenarios are given in Table \ref{tab:prob.T.leq.C.sim}. From this we see that under Scenario 1 the uncensoring probabilities are quite constant over the studied dependence strength and similar for all investigated copula families. For Scenario 2 we observe lower uncensoring probabilities and they decrease as the dependence strength increases. 

\begin{table}[ht]
\centering
\begin{tabular}{rrrr}
  \hline
 Scenario 1: Copula & $\tau=.2$ & $\tau=.5 $& $\tau=.8$ \\ 
  \hline
Frank & 0.41 & 0.40 & 0.39 \\ 
  Clayton & 0.43 & 0.43 & 0.40 \\ 
  Gumbel & 0.41 & 0.40 & 0.39 \\ 
  Gauss & 0.42 & 0.41 & 0.40 \\ 
   \hline
Scenario 2: Copula & $\tau=.2$ & $\tau=.5$ & $\tau=.8$ \\ 
  \hline
Frank & 0.28 & 0.24 & 0.16 \\ 
  Clayton & 0.31 & 0.27 & 0.18 \\ 
  Gumbel & 0.30 & 0.24 & 0.18 \\ 
  Gauss & 0.30 & 0.25 & 0.18 \\ 
   \hline
\end{tabular}
\caption{$P(\Delta=1)$ for all simulation scenarios.}
\label{tab:prob.T.leq.C.sim}
\end{table}

We allow for three different sample sizes $n=200, 500$ and $1000$ and repeated each simulation setting 100 times. We report the average estimate (\texttt{average.estimate}) together with the average bias (\texttt{average.bias}), the standard deviation of the average estimate (\texttt{sd.of.average.estimate}) and the empirical root mean squared error (\texttt{RMSE}) based on the 100 replications.


The results for Scenario 1 are shown in Tables \ref{sim1:frank}, \ref{sim1:clayton},
\ref{sim1:gumbel} and \ref{sim1:gauss} for the Frank, Clayton, Gumbel and Gauss copula, respectively. We see satisfactory performance of the estimation procedure. Only for the Frank copula we see a considerable bias for the small dependence strength of $\tau=.2$. As we expect the average RMSE goes down as the sample size increases. The marginal parameters are well estimated in all cases. 

\begin{table}[ht]
\centering
\begin{footnotesize}
\begin{tabular}{rrrrrrr}
  \hline
 $\tau=.2, n=200$&  $\mu_T$ &  $\log(\sigma_T)$ &  $\mu_C$ &  $\log(\sigma_C)$ &  
 $\mbox{logit}(\tau)$ & $\tau$\\
  \hline
average.estimate & 2.17 & -0.02 & 1.99 & -1.38 & -0.63 & 0.35 \\ 
  sd.of.average.estimate & 0.01 & 0.01 & 0.00 & 0.01 & 0.04 & 0.01 \\ 
  average.bias & -0.02 & -0.02 & -0.01 & 0.01 & 0.76 & 0.15 \\ 
  RMSE & 0.10 & 0.09 & 0.03 & 0.07 & 0.84 & 0.17 \\ 
   \hline
     $n=500$ &  $\mu_T$ &  $\log(\sigma_T)$ &  $\mu_C$ &  $\log(\sigma_C)$ &  $\mbox{logit}(\tau)$ & $\tau$\\
     \hline
  average.estimate & 2.19 & -0.02 & 1.99 & -1.38 & -0.66 & 0.34 \\ 
  sd.of.average.estimate & 0.00 & 0.01 & 0.00 & 0.00 & 0.02 & 0.01 \\ 
  average.bias & -0.01 & -0.02 & -0.01 & 0.01 & 0.72 & 0.14 \\ 
  RMSE & 0.05 & 0.06 & 0.02 & 0.05 & 0.76 & 0.15 \\ 
   \hline
     $n=1000$ &  $\mu_T$ &  $\log(\sigma_T)$ &  $\mu_C$ &  $\log(\sigma_C)$ &  $\mbox{logit}(\tau)$ & $\tau$\\
     \hline
  average.estimate & 2.18 & -0.01 & 1.99 & -1.38 & -0.73 & 0.33 \\ 
  sd.of.average.estimate & 0.00 & 0.00 & 0.00 & 0.00 & 0.02 & 0.00 \\ 
  average.bias & -0.01 & -0.01 & -0.01 & 0.01 & 0.66 & 0.13 \\ 
  RMSE & 0.04 & 0.04 & 0.01 & 0.03 & 0.68 & 0.13 \\ 
    \hline
 $\tau=.5, n=200$&  $\mu_T$ &  $\log(\sigma_T)$ &  $\mu_C$ &  $\log(\sigma_C)$ &  
 $\mbox{logit}(\tau)$ & $\tau$\\
  \hline
  average.estimate & 2.20 & -0.01 & 2.00 & -1.38 & -0.02 & 0.49 \\ 
  sd.of.average.estimate & 0.01 & 0.01 & 0.00 & 0.01 & 0.05 & 0.01 \\ 
  average.bias & -0.00 & -0.01 & 0.00 & 0.00 & -0.02 & -0.00 \\ 
  RMSE & 0.10 & 0.10 & 0.03 & 0.09 & 0.51 & 0.12 \\ 
   \hline
     $n=500$ &  $\mu_T$ &  $\log(\sigma_T)$ &  $\mu_C$ &  $\log(\sigma_C)$ &  $\mbox{logit}(\tau)$ & $\tau$\\
     \hline
  average.estimate & 2.20 & -0.01 & 2.00 & -1.38 & -0.02 & 0.49 \\ 
  sd.of.average.estimate & 0.01 & 0.01 & 0.00 & 0.00 & 0.03 & 0.01 \\ 
  average.bias & 0.00 & -0.01 & -0.00 & 0.00 & -0.02 & -0.00 \\ 
  RMSE & 0.06 & 0.07 & 0.02 & 0.05 & 0.34 & 0.08 \\ 
   \hline
     $n=1000$ &  $\mu_T$ &  $\log(\sigma_T)$ &  $\mu_C$ &  $\log(\sigma_C)$ &  $\mbox{logit}(\tau)$ & $\tau$\\
     \hline
  average.estimate & 2.20 & 0.00 & 2.00 & -1.38 & -0.01 & 0.50 \\ 
  sd.of.average.estimate & 0.00 & 0.00 & 0.00 & 0.00 & 0.02 & 0.00 \\ 
  average.bias & -0.00 & 0.00 & -0.00 & 0.00 & -0.01 & -0.00 \\ 
  RMSE & 0.05 & 0.04 & 0.01 & 0.04 & 0.22 & 0.05 \\ 
    \hline
 $\tau=.8, n=200$&  $\mu_T$ &  $\log(\sigma_T)$ &  $\mu_C$ &  $\log(\sigma_C)$ &  
 $\mbox{logit}(\tau)$ & $\tau$\\
  \hline
  average.estimate & 2.21 & 0.00 & 2.00 & -1.40 & 1.39 & 0.80 \\ 
  sd.of.average.estimate & 0.01 & 0.01 & 0.00 & 0.01 & 0.03 & 0.00 \\ 
  average.bias & 0.00 & 0.00 & 0.00 & -0.01 & 0.01 & -0.00 \\ 
  RMSE & 0.10 & 0.08 & 0.02 & 0.09 & 0.29 & 0.04 \\ 
   \hline
     $n=500$ &  $\mu_T$ &  $\log(\sigma_T)$ &  $\mu_C$ &  $\log(\sigma_C)$ &  $\mbox{logit}(\tau)$ & $\tau$\\
     \hline
  average.estimate & 2.19 & -0.01 & 2.00 & -1.38 & 1.39 & 0.80 \\ 
  sd.of.average.estimate & 0.01 & 0.01 & 0.00 & 0.00 & 0.02 & 0.00 \\ 
  average.bias & -0.01 & -0.01 & -0.00 & 0.00 & 0.00 & -0.00 \\ 
  RMSE & 0.06 & 0.06 & 0.01 & 0.05 & 0.18 & 0.03 \\ 
   \hline
     $n=1000$ &  $\mu_T$ &  $\log(\sigma_T)$ &  $\mu_C$ &  $\log(\sigma_C)$ &  $\mbox{logit}(\tau)$ & $\tau$\\
     \hline
  average.estimate & 2.20 & -0.00 & 2.00 & -1.38 & 1.40 & 0.80 \\ 
  sd.of.average.estimate & 0.00 & 0.00 & 0.00 & 0.00 & 0.01 & 0.00 \\ 
  average.bias & -0.00 & -0.00 & -0.00 & 0.00 & 0.01 & 0.00 \\ 
  RMSE & 0.04 & 0.04 & 0.01 & 0.04 & 0.14 & 0.02 \\[-.1cm]
   \hline
\end{tabular}
\caption{Simulation results for the case of a Frank copula under Scenario 1.}
\label{sim1:frank}
\end{footnotesize}
\end{table}

     \begin{table}[ht]
     \centering
     \begin{footnotesize}
     \begin{tabular}{rrrrrrr}
     \hline
     $\tau=.2, n=200$&  $\mu_T$ &  $\log(\sigma_T)$ &  $\mu_C$ &  $\log(\sigma_C)$ &  
 $\mbox{logit}(\tau)$ & $\tau$\\
     \hline
     average.estimate & 2.20 & -0.03 & 1.99 & -1.37 & -2.01 & 0.23 \\
     sd.of.average.estimate & 0.01 & 0.01 & 0.00 & 0.01 & 0.22 & 0.02 \\
     average.bias & 0.00 & -0.03 & -0.01 & 0.02 & -0.63 & 0.03 \\
     RMSE & 0.11 & 0.10 & 0.05 & 0.10 & 2.29 & 0.18 \\
     \hline
     $n=500$ &  $\mu_T$ &  $\log(\sigma_T)$ &  $\mu_C$ &  $\log(\sigma_C)$ &  $\mbox{logit}(\tau)$ & $\tau$\\
     \hline
     average.estimate      & 2.19 & -0.01 & 2.00 & -1.38 & -1.89 & 0.21 \\
     sd.of.average.estimate      & 0.01 & 0.01 & 0.00 & 0.01 & 0.17 & 0.01 \\
     average.bias      & -0.00 & -0.01 & -0.00 & 0.00 & -0.50 & 0.01 \\
     RMSE      & 0.07 & 0.06 & 0.04 & 0.08 & 1.81 & 0.14 \\
     \hline
     $n=1000$ &  $\mu_T$ &  $\log(\sigma_T)$ &  $\mu_C$ &  $\log(\sigma_C)$ &  $\mbox{logit}(\tau)$ & $\tau$\\
     \hline
     average.estimate      & 2.20 & 0.00 & 2.00 & -1.38 & -1.57 & 0.21 \\
     sd.of.average.estimate      & 0.00 & 0.00 & 0.00 & 0.00 & 0.11 & 0.01 \\
     average.bias      & 0.00 & 0.00 & -0.00 & 0.00 & -0.18 & 0.01 \\
     RMSE      & 0.05 & 0.04 & 0.03 & 0.05 & 1.07 & 0.10 \\
     \hline
     $\tau=.5, n=200$&  $\mu_T$ &  $\log(\sigma_T)$ &  $\mu_C$ &  $\log(\sigma_C)$ &  
 $\mbox{logit}(\tau)$ & $\tau$\\
     \hline
     average.estimate      & 2.20 & -0.00 & 2.00 & -1.40 & -0.14 & 0.49 \\
     sd.of.average.estimate      & 0.01 & 0.01 & 0.00 & 0.01 & 0.10 & 0.01 \\
     average.bias      & 0.00 & -0.00 & 0.00 & -0.01 & -0.14 & -0.01 \\
     RMSE      & 0.10 & 0.09 & 0.04 & 0.11 & 1.04 & 0.15 \\
     \hline
     $n=500$ &  $\mu_T$ &  $\log(\sigma_T)$ &  $\mu_C$ &  $\log(\sigma_C)$ &  $\mbox{logit}(\tau)$ & $\tau$\\
     \hline
     average.estimate      & 2.21 & -0.00 & 2.00 & -1.39 & -0.02 & 0.49 \\
     sd.of.average.estimate      & 0.01 & 0.01 & 0.00 & 0.01 & 0.04 & 0.01 \\
     average.bias      & 0.01 & -0.00 & 0.00 & -0.01 & -0.02 & -0.00 \\
     RMSE      & 0.07 & 0.06 & 0.02 & 0.06 & 0.35 & 0.08 \\
     \hline
     $n=1000$ &  $\mu_T$ &  $\log(\sigma_T)$ &  $\mu_C$ &  $\log(\sigma_C)$ &  $\mbox{logit}(\tau)$ & $\tau$\\
     \hline
     average.estimate      & 2.20 & -0.01 & 2.00 & -1.38 & 0.02 & 0.51 \\
     sd.of.average.estimate      & 0.00 & 0.00 & 0.00 & 0.00 & 0.02 & 0.00 \\
     average.bias      & 0.00 & -0.01 & -0.00 & 0.00 & 0.02 & 0.01 \\
     RMSE      & 0.04 & 0.04 & 0.02 & 0.04 & 0.22 & 0.06 \\
     \hline
     $\tau=.8, n=200$&  $\mu_T$ &  $\log(\sigma_T)$ &  $\mu_C$ &  $\log(\sigma_C)$ &  
 $\mbox{logit}(\tau)$ & $\tau$\\
     \hline
     average.estimate      & 2.20 & -0.01 & 2.00 & -1.37 & 1.45 & 0.81 \\
     sd.of.average.estimate      & 0.01 & 0.01 & 0.00 & 0.01 & 0.03 & 0.00 \\
     average.bias      & -0.00 & -0.01 & 0.00 & 0.01 & 0.07 & 0.01 \\
     RMSE      & 0.09 & 0.08 & 0.02 & 0.07 & 0.33 & 0.05 \\
     \hline
     $n=500$ &  $\mu_T$ &  $\log(\sigma_T)$ &  $\mu_C$ &  $\log(\sigma_C)$ &  $\mbox{logit}(\tau)$ & $\tau$\\
     \hline
     average.estimate      & 2.20 & -0.01 & 2.00 & -1.39 & 1.41 & 0.80 \\
     sd.of.average.estimate      & 0.01 & 0.01 & 0.00 & 0.00 & 0.02 & 0.00 \\
     average.bias      & 0.00 & -0.01 & 0.00 & 0.00 & 0.02 & 0.00 \\
     RMSE      & 0.07 & 0.06 & 0.01 & 0.05 & 0.20 & 0.03 \\
     \hline
     $n=1000$ &  $\mu_T$ &  $\log(\sigma_T)$ &  $\mu_C$ &  $\log(\sigma_C)$ &  $\mbox{logit}(\tau)$ & $\tau$\\
     \hline
     average.estimate      & 2.20 & -0.00 & 2.00 & -1.39 & 1.40 & 0.80 \\
     sd.of.average.estimate      & 0.00 & 0.00 & 0.00 & 0.00 & 0.01 & 0.00 \\
     average.bias      & -0.00 & -0.00 & 0.00 & -0.00 & 0.01 & 0.00 \\
     RMSE      & 0.04 & 0.04 & 0.01 & 0.04 & 0.15 & 0.02 \\[-.1cm]
     \hline
     \end{tabular}
     \end{footnotesize}
     \caption{Simulation results for the case of a Clayton copula under Scenario 1.}
     \label{sim1:clayton}
     \end{table}

     \begin{table}[ht]
     \centering
     \begin{footnotesize}
     \begin{tabular}{rrrrrrr}
     \hline
     $\tau=.2, n=200$&  $\mu_T$ &  $\log(\sigma_T)$ &  $\mu_C$ &  $\log(\sigma_C)$ &  
 $\mbox{logit}(\tau)$ & $\tau$\\
   \hline
     average.estimate & 2.19 & -0.01 & 2.00 & -1.37 & -1.92 & 0.21 \\
     sd.of.average.estimate & 0.01 & 0.01 & 0.00 & 0.01 & 0.19 & 0.01 \\
     average.bias & -0.00 & -0.01 & 0.00 & 0.01 & -0.53 & 0.01 \\
     RMSE & 0.10 & 0.10 & 0.04 & 0.07 & 2.00 & 0.14 \\
     \hline
     $n=500$ &  $\mu_T$ &  $\log(\sigma_T)$ &  $\mu_C$ &  $\log(\sigma_C)$ &  $\mbox{logit}(\tau)$ & $\tau$\\
     \hline
     average.estimate      & 2.20 & -0.00 & 2.00 & -1.39 & -1.58 & 0.21 \\
     sd.of.average.estimate      & 0.01 & 0.00 & 0.00 & 0.00 & 0.11 & 0.01 \\
     average.bias      & -0.00 & -0.00 & 0.00 & 0.00 & -0.19 & 0.01 \\
     RMSE      & 0.06 & 0.05 & 0.02 & 0.04 & 1.14 & 0.09 \\
     \hline
     $n=1000$ &  $\mu_T$ &  $\log(\sigma_T)$ &  $\mu_C$ &  $\log(\sigma_C)$ &  $\mbox{logit}(\tau)$ & $\tau$\\
     \hline
     average.estimate      & 2.20 & -0.00 & 2.00 & -1.38 & -1.39 & 0.21 \\
     sd.of.average.estimate      & 0.00 & 0.00 & 0.00 & 0.00 & 0.05 & 0.01 \\
     average.bias      & 0.00 & -0.00 & -0.00 & 0.00 & -0.00 & 0.01 \\
     RMSE      & 0.05 & 0.05 & 0.02 & 0.03 & 0.50 & 0.07 \\
     \hline
      $\tau=.5, n=200$&  $\mu_T$ &  $\log(\sigma_T)$ &  $\mu_C$ &  $\log(\sigma_C)$ &  
 $\mbox{logit}(\tau)$ & $\tau$\\
     \hline
     average.estimate      & 2.20 & -0.01 & 2.00 & -1.38 & 0.03 & 0.51 \\
     sd.of.average.estimate      & 0.01 & 0.01 & 0.00 & 0.01 & 0.06 & 0.01 \\
     average.bias      & -0.00 & -0.01 & -0.00 & 0.01 & 0.03 & 0.01 \\
     RMSE      & 0.11 & 0.09 & 0.03 & 0.08 & 0.57 & 0.13 \\
     \hline
     $n=500$ &  $\mu_T$ &  $\log(\sigma_T)$ &  $\mu_C$ &  $\log(\sigma_C)$ &  $\mbox{logit}(\tau)$ & $\tau$\\
     \hline
     average.estimate      & 2.20 & -0.01 & 2.00 & -1.39 & -0.03 & 0.49 \\
     sd.of.average.estimate      & 0.01 & 0.01 & 0.00 & 0.01 & 0.04 & 0.01 \\
     average.bias      & 0.00 & -0.01 & 0.00 & -0.01 & -0.03 & -0.01 \\
     RMSE      & 0.07 & 0.07 & 0.02 & 0.06 & 0.37 & 0.09 \\
     \hline
     $n=1000$ &  $\mu_T$ &  $\log(\sigma_T)$ &  $\mu_C$ &  $\log(\sigma_C)$ &  $\mbox{logit}(\tau)$ & $\tau$\\
     \hline
     average.estimate      & 2.21 & 0.00 & 2.00 & -1.38 & 0.01 & 0.50 \\
     sd.of.average.estimate      & 0.00 & 0.00 & 0.00 & 0.00 & 0.02 & 0.01 \\
     average.bias      & 0.01 & 0.00 & 0.00 & 0.00 & 0.01 & 0.00 \\
     RMSE      & 0.05 & 0.04 & 0.01 & 0.04 & 0.22 & 0.06 \\
     \hline
      $\tau=.8, n=200$&  $\mu_T$ &  $\log(\sigma_T)$ &  $\mu_C$ &  $\log(\sigma_C)$ &  
 $\mbox{logit}(\tau)$ & $\tau$\\
     \hline
     average.estimate      & 2.19 & -0.02 & 2.00 & -1.40 & 1.42 & 0.80 \\
     sd.of.average.estimate      & 0.01 & 0.01 & 0.00 & 0.01 & 0.03 & 0.00 \\
     average.bias      & -0.01 & -0.02 & -0.00 & -0.01 & 0.04 & 0.00 \\
     RMSE      & 0.10 & 0.10 & 0.02 & 0.08 & 0.32 & 0.05 \\
     \hline
     $n=500$ &  $\mu_T$ &  $\log(\sigma_T)$ &  $\mu_C$ &  $\log(\sigma_C)$ &  $\mbox{logit}(\tau)$ & $\tau$\\
     \hline
     average.estimate      & 2.20 & 0.00 & 2.00 & -1.39 & 1.40 & 0.80 \\
     sd.of.average.estimate      & 0.01 & 0.00 & 0.00 & 0.00 & 0.02 & 0.00 \\
     average.bias      & -0.00 & 0.00 & -0.00 & -0.01 & 0.01 & 0.00 \\
     RMSE      & 0.06 & 0.05 & 0.02 & 0.05 & 0.19 & 0.03 \\
     \hline
     $n=1000$ &  $\mu_T$ &  $\log(\sigma_T)$ &  $\mu_C$ &  $\log(\sigma_C)$ &  $\mbox{logit}(\tau)$ & $\tau$\\
     \hline
     average.estimate      & 2.20 & -0.00 & 2.00 & -1.39 & 1.39 & 0.80 \\
     sd.of.average.estimate      & 0.00 & 0.00 & 0.00 & 0.00 & 0.01 & 0.00 \\
     average.bias      & 0.00 & -0.00 & 0.00 & 0.00 & 0.00 & -0.00 \\
     RMSE      & 0.04 & 0.04 & 0.01 & 0.04 & 0.13 & 0.02 \\[-.1cm]
     \hline
     \end{tabular}
     \end{footnotesize}
     \caption{Simulation results for the case of a Gumbel copula under Scenario 1.}
     \label{sim1:gumbel}
     \end{table}

\begin{table}[ht]
\centering
\begin{footnotesize}
\begin{tabular}{rrrrrrr}
 \hline
     $\tau=.2, n=200$&  $\mu_T$ &  $\log(\sigma_T)$ &  $\mu_C$ &  $\log(\sigma_C)$ &  
 $\mbox{logit}(\tau)$ & $\tau$\\
   \hline
  average.estimate & 2.46 & -0.05 & 1.98 & -0.68 & -2.08 & 0.24 \\ 
  sd.of.average.estimate & 0.03 & 0.02 & 0.01 & 0.01 & 0.23 & 0.02 \\ 
  average.bias & -0.04 & -0.05 & -0.02 & 0.02 & -0.69 & 0.04 \\ 
  RMSE & 0.31 & 0.17 & 0.13 & 0.11 & 2.37 & 0.20 \\ 
   \hline
     $n=500$ &  $\mu_T$ &  $\log(\sigma_T)$ &  $\mu_C$ &  $\log(\sigma_C)$ &  $\mbox{logit}(\tau)$ & $\tau$\\
     \hline
  average.estimate & 2.19 & -0.01 & 1.99 & -1.36 & -2.08 & 0.19 \\ 
  sd.of.average.estimate & 0.02 & 0.01 & 0.01 & 0.02 & 0.18 & 0.01 \\ 
  average.bias & -0.01 & -0.01 & -0.01 & 0.03 & -0.70 & -0.01 \\ 
  RMSE & 0.22 & 0.08 & 0.11 & 0.17 & 1.94 & 0.12 \\ 
  \hline
     $n=1000$ &  $\mu_T$ &  $\log(\sigma_T)$ &  $\mu_C$ &  $\log(\sigma_C)$ &  $\mbox{logit}(\tau)$ & $\tau$\\
     \hline
  average.estimate & 2.18 & -0.00 & 1.99 & -1.36 & -1.56 & 0.21 \\ 
  sd.of.average.estimate & 0.02 & 0.01 & 0.01 & 0.02 & 0.11 & 0.01 \\ 
  average.bias & -0.02 & -0.00 & -0.01 & 0.02 & -0.17 & 0.01 \\ 
  RMSE & 0.21 & 0.06 & 0.11 & 0.17 & 1.12 & 0.10 \\ 
   \hline
     $\tau=.5, n=200$&  $\mu_T$ &  $\log(\sigma_T)$ &  $\mu_C$ &  $\log(\sigma_C)$ &  
 $\mbox{logit}(\tau)$ & $\tau$\\
   \hline
  average.estimate & 2.19 & -0.02 & 1.99 & -1.38 & -0.19 & 0.48 \\ 
  sd.of.average.estimate & 0.02 & 0.01 & 0.01 & 0.02 & 0.11 & 0.01 \\ 
  average.bias & -0.01 & -0.02 & -0.01 & 0.01 & -0.19 & -0.02 \\ 
  RMSE & 0.24 & 0.11 & 0.11 & 0.19 & 1.11 & 0.15 \\ 
   \hline
     $n=500$ &  $\mu_T$ &  $\log(\sigma_T)$ &  $\mu_C$ &  $\log(\sigma_C)$ &  $\mbox{logit}(\tau)$ & $\tau$\\
     \hline
  average.estimate & 2.18 & -0.01 & 1.99 & -1.38 & -0.14 & 0.48 \\ 
  sd.of.average.estimate & 0.02 & 0.01 & 0.01 & 0.02 & 0.09 & 0.01 \\ 
  average.bias & -0.02 & -0.01 & -0.01 & 0.01 & -0.14 & -0.02 \\ 
  RMSE & 0.21 & 0.07 & 0.11 & 0.18 & 0.94 & 0.10 \\ 
 \hline
     $n=1000$ &  $\mu_T$ &  $\log(\sigma_T)$ &  $\mu_C$ &  $\log(\sigma_C)$ &  $\mbox{logit}(\tau)$ & $\tau$\\
     \hline
  average.estimate & 2.18 & -0.00 & 1.99 & -1.37 & -0.09 & 0.49 \\ 
  sd.of.average.estimate & 0.02 & 0.01 & 0.01 & 0.02 & 0.09 & 0.01 \\ 
  average.bias & -0.02 & -0.00 & -0.01 & 0.02 & -0.09 & -0.01 \\ 
  RMSE & 0.21 & 0.06 & 0.11 & 0.17 & 0.89 & 0.08 \\ 
   \hline
     $\tau=.8, n=200$&  $\mu_T$ &  $\log(\sigma_T)$ &  $\mu_C$ &  $\log(\sigma_C)$ &  
 $\mbox{logit}(\tau)$ & $\tau$\\
   \hline
  average.estimate & 2.18 & -0.01 & 1.99 & -1.38 & 1.30 & 0.79 \\ 
  sd.of.average.estimate & 0.02 & 0.01 & 0.01 & 0.02 & 0.10 & 0.01 \\ 
  average.bias & -0.02 & -0.01 & -0.01 & 0.00 & -0.09 & -0.01 \\ 
  RMSE & 0.24 & 0.11 & 0.11 & 0.18 & 1.04 & 0.09 \\ 
  \hline
     $n=500$ &  $\mu_T$ &  $\log(\sigma_T)$ &  $\mu_C$ &  $\log(\sigma_C)$ &  $\mbox{logit}(\tau)$ & $\tau$\\
     \hline
  average.estimate & 2.17 & -0.01 & 1.99 & -1.38 & 1.28 & 0.79 \\ 
  sd.of.average.estimate & 0.02 & 0.01 & 0.01 & 0.02 & 0.10 & 0.01 \\ 
  average.bias & -0.03 & -0.01 & -0.01 & 0.01 & -0.10 & -0.01 \\ 
  RMSE & 0.22 & 0.08 & 0.11 & 0.17 & 1.01 & 0.08 \\ 
  \hline
     $n=1000$ &  $\mu_T$ &  $\log(\sigma_T)$ &  $\mu_C$ &  $\log(\sigma_C)$ &  $\mbox{logit}(\tau)$ & $\tau$\\
     \hline
  average.estimate & 2.18 & -0.00 & 1.99 & -1.38 & 1.29 & 0.79 \\ 
  sd.of.average.estimate & 0.02 & 0.01 & 0.01 & 0.02 & 0.10 & 0.01 \\ 
  average.bias & -0.02 & -0.00 & -0.01 & 0.01 & -0.10 & -0.01 \\ 
  RMSE & 0.21 & 0.06 & 0.11 & 0.17 & 1.00 & 0.08 \\[-.1cm]
   \hline
\end{tabular}
\end{footnotesize}
\caption{Simulation results for the case of a Gauss copula under Scenario 1.}
\label{sim1:gauss}
\end{table}

The results from Scenario 2 are shown in the supplement as  Tables \ref{sim2:frank}, \ref{sim2:clayton},
\ref{sim2:gumbel} and \ref{sim2:gauss} for the Frank, Clayton, Gumbel and Gauss copula, respectively. Compared to Scenario 1 the positive bias for the Frank copula is even higher under Scenario 2 for low dependence. In the other copula cases we also see that there is a slightly higher bias in the low dependence case compared to Scenario 1. However for stronger dependences the performance is satisfactory even for sample sizes as low as 200.  

\clearpage

\section{Illustration} \label{sect6}

We now illustrate the dependent censoring model specified in \eqref{model1}-\eqref{model2} for pancreas cancer  from the Surveillance, Epidemiology, and End Results (SEER) database (see \url{https://seer.cancer.gov/data-software/}).
More specifically we use the monthly survival times of blacks with localized
pancreas cancer between 2000 and 2015. We exclude all patients
with 0 survival times. 
This leaves us with data on 1549 patients with localized pancreas cancer, of which 777 died of pancreas cancer and 772 patients were still alive or died of other causes. In our analysis we view the 772 patients as censored observations.   When patients are censored because they died from another disease, their censoring time is likely related to their (unobserved) survival time, since many diseases share common risk factors (like stress, eating habits, physical condition, etc.). 

We now fit model \eqref{model1}-\eqref{model2} for the independence, Frank, Clayton, Gumbel and Gauss copula with log-normal or Weibull margins to this data set. 
The resulting parameter estimates are given in Tables \ref{tab:estimates-lognormal} and \ref{tab:estimates-weibull}, respectively. \tcr For the Weibull margins we have the shape parameter $a>0$ and scale parameter $b>0$ to estimate, where we use now the parametrization of the Weibull density used in R (note that the parameters $\lambda$ and $\rho$ used in (\ref{Weib}) can be expressed in terms of $a$ and $b$ via the formulas $\lambda=b^{-a}$ and $\rho=a$). \tcb
We estimate the unconstrained parameters $\log(a)$ and $\log(b)$. Similarly the unconstrained parameters $\mu$ and $\log(\sigma)$ for the log-normal margins are estimated. All parameters are estimated using maximum likelihood. Standard error estimates are based on using 100 bootstrap samples with replacement. In view of the asymptotic theory given in Theorem  \ref{asnorm} we expect the bootstrap procedure to give valid standard error estimates.

We now compare the different marginal and copula specifications.
Since the model complexities of these single parametric copula models are the same, considering the maximized log-likelihood is sufficient.
From the estimation results we see that the fitted log likelihoods for the copula models with log-normal margins are higher than the ones with Weibull margins. The increase ranges from 13.5 for the Gaussian copula model to 100.8 for the Frank copula models. These models are non-nested, but all have the same number of parameters. Therefore there is a preference for the models with log-normal margins. 

With regard to the copula specification  the models with the Gumbel copula have the highest log likelihood. 
Comparing the dependent censoring models to the independent censoring one, we see a drop of more than 100 for both marginal models for the Gumbel specification. Therefore a likelihood ratio test is very much in favor of the dependent censoring model using the Gumbel copula. This indicates that upper tail dependence is present between the survival times and the censoring times. This means that long survival or censoring times occur much more often together than short times. This is also supported by the worse performance of the Clayton copula based model, which allows for lower tail dependence and no upper tail dependence. The symmetric but no tail dependent Gaussian and Frank copula models perform better than the Clayton copula but worse than the Gumbel copula. The worse performance of the Frank over the Gauss copula can be attributed to the fact that the Frank copula has even lighter joint tails than the Gaussian copula.

The fitted marginal probabilities $P(\Delta=1)=P(T \leq C)$ are given in Table \ref{tab:prob.T.leq.C}. From this table we see that all model specifications give fitted marginal probabilities that  are close to the  empirical values.

\begin{figure}[h!]
\centering
\includegraphics[scale=.52]{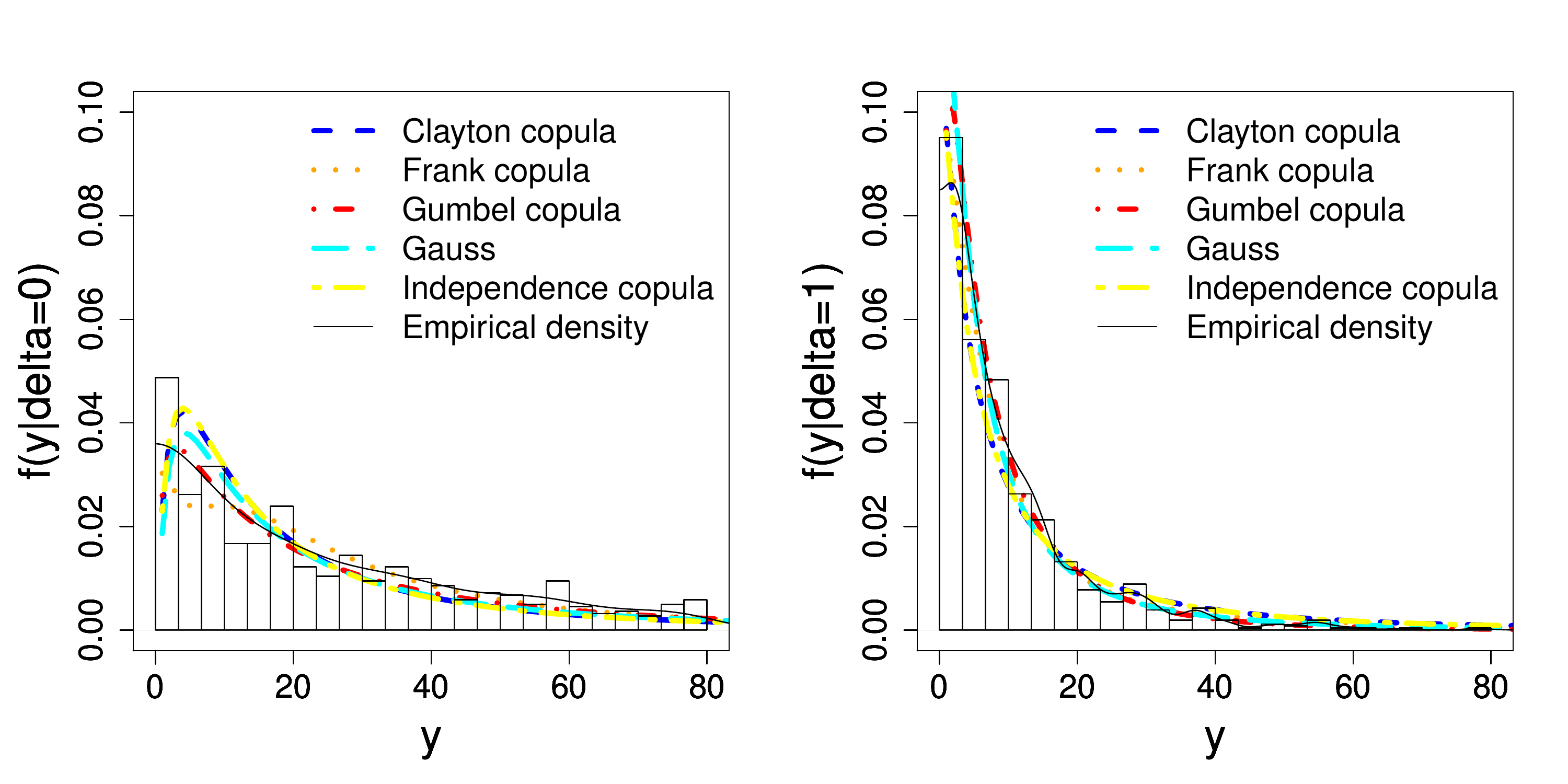}
\includegraphics[scale=.5]{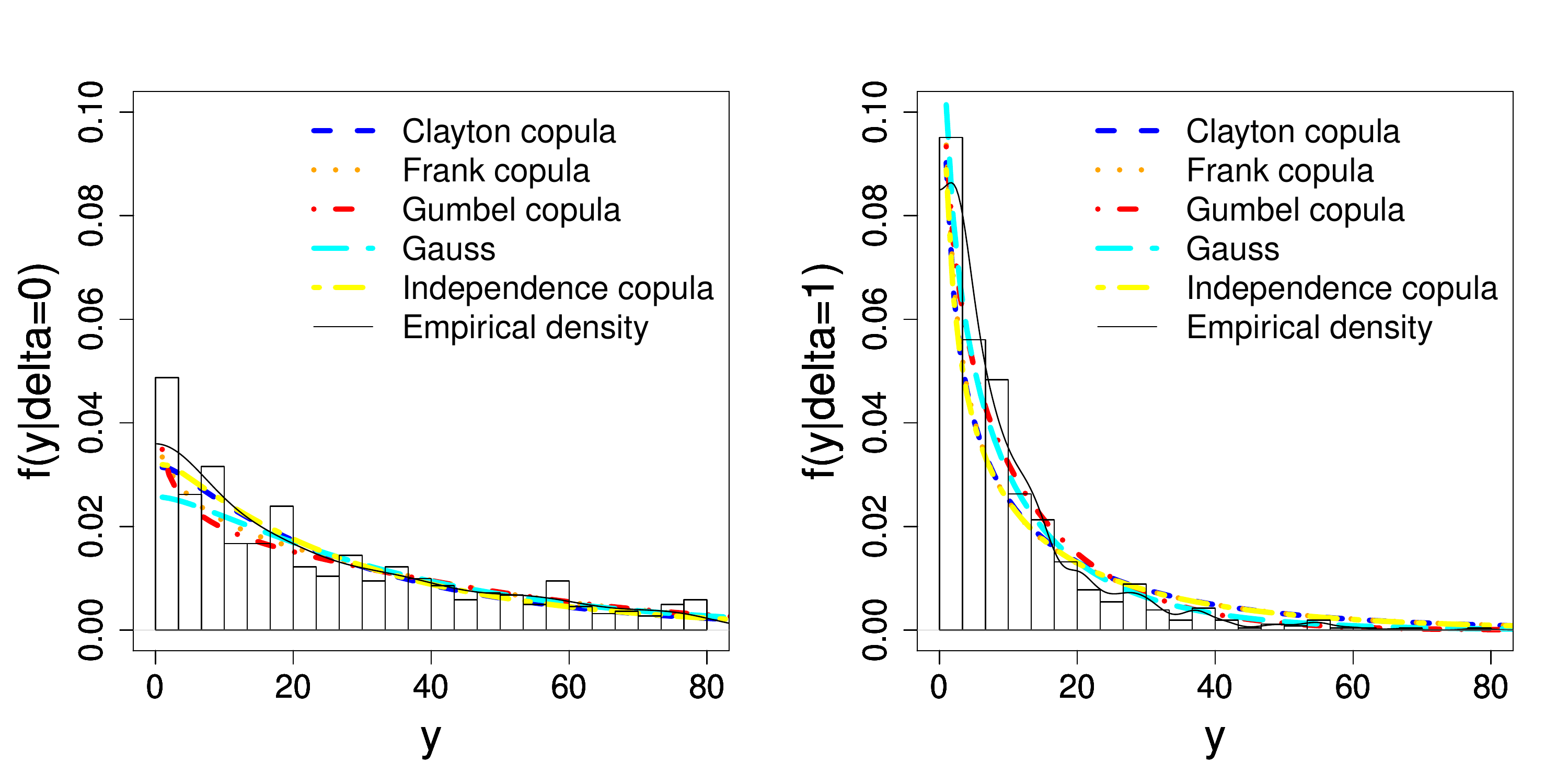}
\caption{Fitted density
	 of $Y$ given $\Delta=0$ (left panel) and of $Y$ given $\Delta=1$ (right panel), respectively, for the  SEER data with the independence, Frank, Clayton, Gumbel and Gauss copula and with log-normal margins (top row)  and Weibull margins (bottom row) together with a histogram and empirical density of the observed data.}
\label{fig:fitdenY}
\end{figure}

To further assess the fit we look at the fitted densities of $Y$ given $\Delta=0$ and $Y$ given $\Delta=1$ for both marginal specifications. These fitted densities are given in Figure \ref{fig:fitdenY}. We see that the fitted distribution of the censored observations have a heavier tail compared to the uncensored observations for both marginal specifications. We further note that the Gumbel copula based model is a bit closer to the empirical density for both censored and uncensored observations.

\begin{figure}[h!]
	\centering
	\includegraphics[scale=.45]{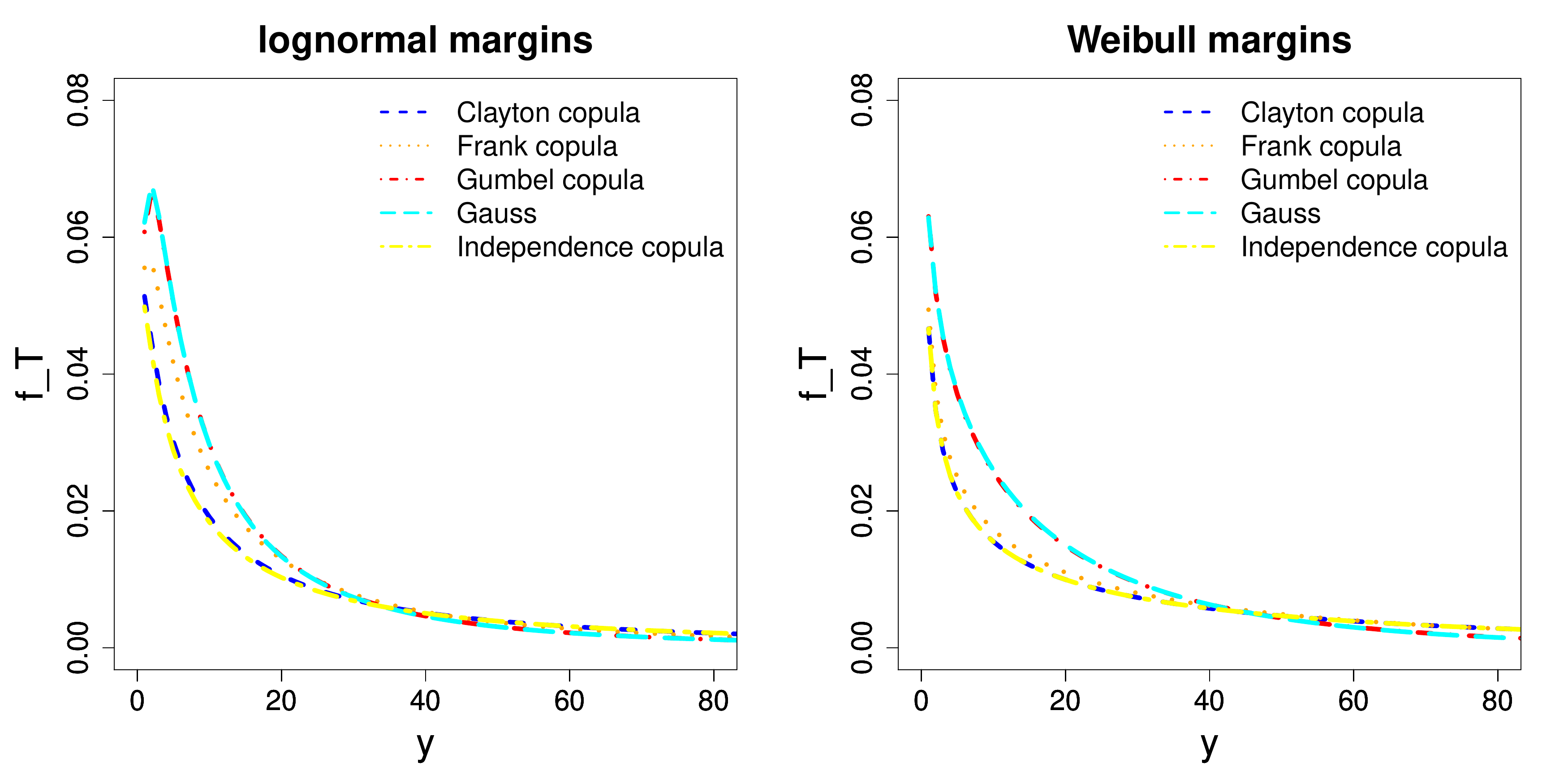}\\
	\includegraphics[scale=.45]{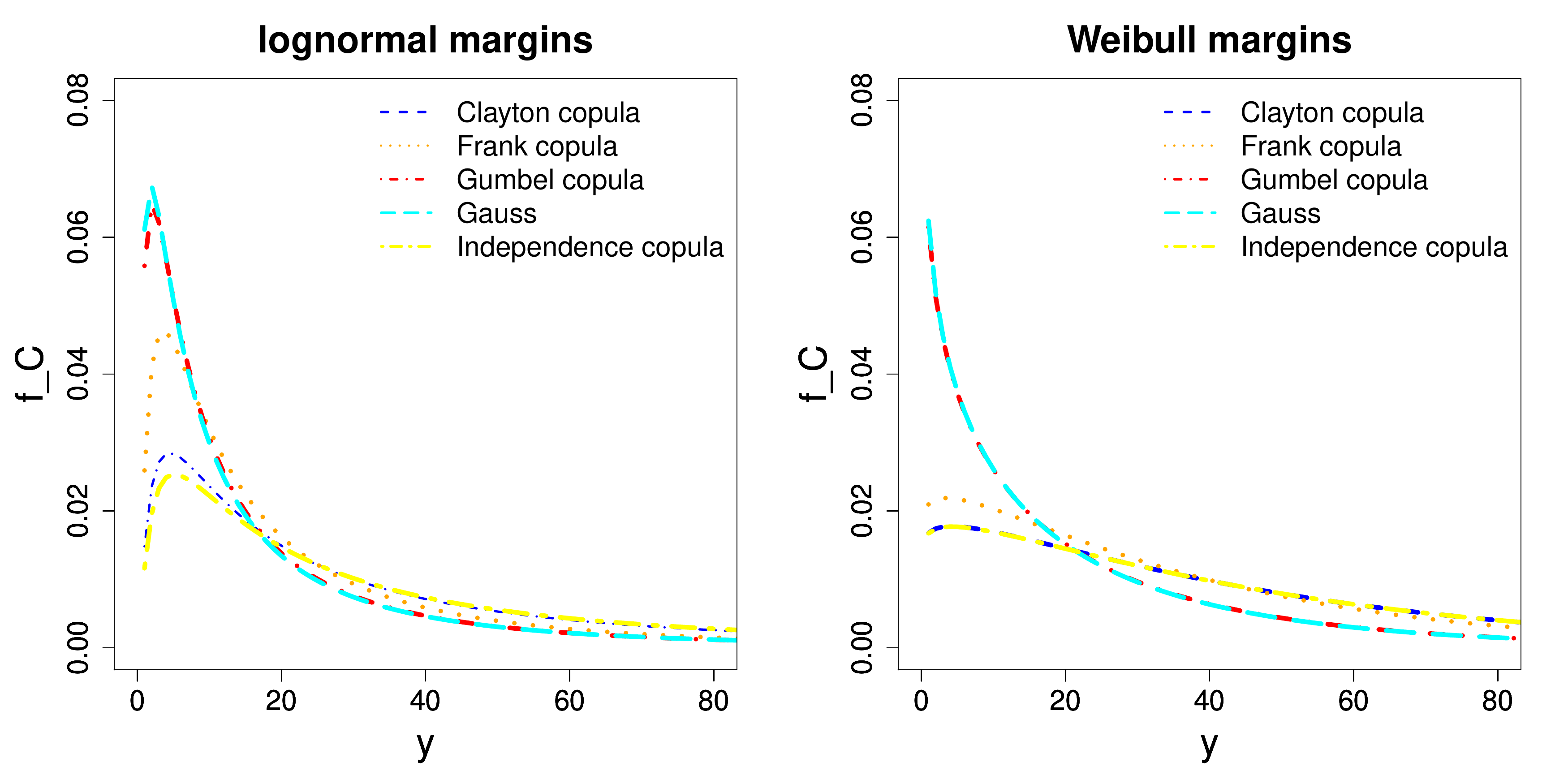}
	\caption{Fitted density of the survival times (top row) and of the censoring times  (bottom row) with log-normal margins
		(left panel) and Weibull margins (right panel), respectively, for the  SEER data with the independence, Frank, Clayton, Gumbel and Gauss copula.}
	\label{fig:fitTandC}
\end{figure}

Finally we give the fitted density of the survival times and censoring times under all copula and marginal specifications
in Figure \ref{fig:fitTandC}. The effect of the copula family is visible for both the survival and censoring times. In particular the copula choice has more influence for the log-normal marginal specification. Further the Gumbel and Gauss copula give very similar fits regardless of the studied marginal specification.

\begin{table}[ht]
\centering
\begin{tabular}{lcccccccc}
  \hline
Copula & $\mu_T$ &  $\log(\sigma_T)$ &  $\mu_C$ &  $\log(\sigma_C)$ &  $\mbox{logit}(\tau)$ & $\theta$ & $\tau$ & log-lik\\
  \hline
  
indep & 3.25 & 0.66 & 3.34 & 0.27 &- & - & - & -7312.37 \\
 (stderr) & (0.06) & (0.02) & (0.04) & (0.02) & - & - & - & (50.04) \\
 Frank & 2.66 & 0.41 & 2.68 & 0.17 & 1.36 & 17.81 & 0.80 & -7231.41 \\
(stderr) & (0.04) & (0.02) & (0.04) & (0.02) & (0.17) & (2.90) & (0.03) & (57.56) \\
Clayton & 3.16 & 0.63 & 3.23 & 0.27 & -2.00 & 0.27 & 0.12 & -7309.74 \\
(stderr) & (0.06) & (0.02) & (0.03) & (0.02) & (0.06) & (0.02) & (0.01) & (43.23) \\
Gumbel & 2.39 & 0.29 & 2.40 & 0.26 & 3.41 & 31.14 & 0.97 & -7194.61 \\
(stderr) & (0.03) & (0.01) & (0.04) & (0.01) & (0.28) & (9.30) & (0.01) & (48.82) \\
Gauss & 2.37 & 0.29 & 2.37 & 0.29 & 4.76 & 1.00 & 0.99 & -7237.34 \\
(stderr) & (0.04) & (0.01) & (0.04) & (0.01) & (0.07) & (0.00) & (0.00) & (55.29) \\

 \hline
\end{tabular}

\caption{Estimated marginal and copula parameters together with the fitted log-likelihood for the SEER data using model (\ref{model1})-(\ref{model2}) with the independence, Frank, Clayton, Gumbel and Gauss copula and log-normal margins, respectively (standard errors are based on 100 bootstrap replications).}
\label{tab:estimates-lognormal}
\end{table}

\begin{table}[ht]
	\centering
	\begin{tabular}{lcccccccc}
		\hline
		Copula & $\log(a_T)$  &  $\log(b_T)$ & $\log(a_C)$  &  $\log(b_C)$ &  $\mbox{logit}(\tau)$ & $\theta$ & $\tau$ & log-lik\\
		\hline
	indep & -0.45 & 3.98 & 0.08 & 3.84 & - & - & - & -7343.55 \\
(stderr) & (0.02) & (0.06) & (0.02) & (0.03) &- & - & - & (37.38) \\
	Frank & -0.40 & 3.77 & 0.08 & 3.63 & -0.83 & 3.00 & 0.30 & -7332.18 \\
	(stderr) & (0.02) & (0.10) & (0.04) & (0.08) & (0.51) & (5.35) & (0.10) & (47.49) \\
	Clayton & -0.45 & 3.98 & 0.08 & 3.84 & -5.92 & 0.00 & 0.00 & -7343.66 \\
	(stderr) & (0.02) & (0.07) & (0.02) & (0.03) & (0.86) & (0.00) & (0.00) & (47.66) \\
	Gumbel & -0.21 & 3.05 & -0.19 & 3.04 & 3.89 & 50.09 & 0.98 & -7241.48 \\
	(stderr) & (0.01) & (0.03) & (0.01) & (0.03) & (0.04) & (1.89) & (0.00) & (47.25) \\
	Gauss & -0.20 & 3.03 & -0.19 & 3.03 & 5.05 & 1.00 & 0.99 & -7250.89 \\
	(stderr) & (0.01) & (0.03) & (0.02) & (0.03) & (0.33) & (0.00) & (0.00) & (42.44) \\

		\hline
	\end{tabular}
	
	\caption{Estimated marginal and copula parameters together with the fitted log-likelihood for the SEER data using model (\ref{model1})-(\ref{model2}) with the independence, Frank, Clayton, Gumbel and Gauss copula and Weibull margins, respectively (standard errors are based on 100 bootstrap replications).}
	\label{tab:estimates-weibull}
\end{table}

\begin{table}[ht]
	\centering
	\begin{tabular}{rrrrrrr}
		\hline
		Margin  & Empirical & Independence &  Frank & Clayton & Gumbel & Gauss \\ 
		\hline
		log-normal &  0.5016 & 0.5160 &  0.5233 & 0.5180 &0.5108 & 0.4973\\
		Weibull & 0.5016 & 0.5084 & 0.5022 &  0.5083 &  0.4935& 0.4932 \\
		\hline
	\end{tabular}
	\caption{Empirical and fitted probability $P(T\leq C)$ for the SEER data with the independence, Frank, Clayton, Gumbel and Gauss copula with log-normal or Weibull margins, respectively.}
	\label{tab:prob.T.leq.C}
\end{table}

\clearpage 

\section{Discussion and future research} \label{sect7}

In this paper we modelled the bivariate distribution of a survival and a censoring time by means of a parametric copula model and parametric margins.  We developed sufficient conditions under which this model is identifiable, and showed that these conditions are satisfied for a wide range of model specifications.   We also studied the estimation of the model and the finite sample performance of the proposed estimators.   

The paper is the first to propose a copula model without assuming that the association parameter of the copula is fully known. This is the main advantage of this paper with respect to existing copula models.   In order to focus on this major gain, we kept the setting of this paper rather simple.  The model can however be extended in many ways.   First, the model can be made more flexible by considering semiparametric or nonparametric margins.  The question is then however whether the identifiability of the model can be guaranteed.    Second, covariates can be added to the model.  The most simple case is that of a fully parametric regression model for the survival and censoring time, but other models like semiparametric Cox proportional hazards or accelerated failure time models will be worth studying as well.    Third, more general survival models, like competing risks, cure models and truncation in combination with dependent censoring, will also be very promising.  Several of the above extensions are currently under preparation.  Finally, we like to mention that we focused here on regular copulas, but we could as well consider survival copulas, or any other rotation of the copula family.  This will not change the essential ideas of the model and its properties.   

\section*{Acknowledgments} 
The authors like to thank Johanna Neslehova for very helpful discussions.  

\bigskip

\bibliographystyle{dcu} 
\bibliography{references}

\section*{Appendix} \label{app}

\setcounter{section}{1}
\renewcommand\theequation{\Alph{section}.\arabic{equation}}

\noindent 
{\bf Derivation of Equations (\ref{resultsYdelta}).}
Differentiating (\ref{model1bis}) we can express the joint density of $(T,C)$ as
\begin{align} 
f_{T,C}(t,c) = c(F_T(t),F_C(c))f_T(t)f_C(c) \label{denTC},
\end{align}
where $c$ denotes the copula density. We are interested in determining the conditional distribution of $T$ given $C$ and vice versa. From (\ref{denTC}) it is straightforward to see that the conditional densities are
\begin{align} 
f_{T|C}(t|c) = c(F_T(t),F_C(c))f_T(t) \label{denTgivenC} \\
f_{C|T}(c|T) = c(F_T(t),F_C(c))f_C(c) \label{denCgivenT}
\end{align}
and the conditional distribution function of $T$ given $C=c$ can be derived as
\begin{eqnarray*} 
F_{T|C}(t|c) &  =  & \int_0^t c(F_T(t^*),F_C(c))f_T(t^*) dt^*  \\
  & = &  \int_0^t \frac{\partial ^2}{\partial u \partial v} \C(u,v)|_{u=F_T(t^*),v=F_C(c)} \frac{d F_T(t)}{d t}|_{t=t^*} dt^*\\
  & = & \frac{\partial}{\partial v} \C(u,v)|_{u=F_T(t),v=F_C(c)}
  = h_{T|C}(F_T(t)|F_C(c)).
\end{eqnarray*}
Similarly we have $F_{C|T}(c|t)=h_{C|T}(F_C(c)|F_T(t))$. 
We now derive the marginal distribution of $Y=\min(T,C)$: 
\begin{eqnarray*}
    F_Y(y)& = & 1 - P(Y>y) = 1- P(T>y, C>y)\\
    & = & 1- \big\{1- F_C(y)-F_T(y) + \C(F_T(y),F_C(y))\big\} \\
    & = & F_C(y) + F_T(y) - \C(F_T(y),F_C(y)).
\end{eqnarray*}
Finally we derive the expressions for the joint mixed density $f_{Y,\Delta}$ by noting that 
\begin{align*}
F_{Y,\Delta}(y,1) &= P(T \le y,T \le C) = \int_0^y P(C \ge t|T=t) f_T(t) \, dt \\
&= \int_0^y \{1-h_{C|T}(F_C(t)|F_T(t)) f_T(t) \, dt,
\end{align*}
and hence $f_{Y,\Delta}(y,1) = \{1-h_{C|T}(F_C(y)|F_T(y))\} f_T(y)$.  
Similarly, we get $f_{Y,\Delta}(y,0) = \{1-h_{T|C}(F_T(y)|F_C(y))\} f_C(y)$. \hfill $\Box$
\vspace*{.5cm}

\noindent
{\bf Proof of Theorem \ref{ident}.}  Recall from (\ref{resultsYdelta}) that 
\begin{align*}
P(Y \le t, \Delta=1) & = P(T \le t, T \le C) = \int_0^t P(C \ge y | T=y) f_T(y) \, dy.
\end{align*}
Hence, 
$$ f_{Y,\Delta}(t,1) = (1-F_{C|T}(t|t)) f_T(t) = \big(1-h_{C|T}(F_C(t)|F_T(t))\big) f_T(t). $$
From condition (C2b) we know that $\lim_{t \rightarrow a} h_{C|T}(F_C(t)|F_T(t)) = 0$ for $a=0$ or $a=\infty$.   Hence, 
$$ \lim_{t \rightarrow a} f_{Y,\Delta}(t,1) = \lim_{t \rightarrow a} f_T(t). $$
Suppose now that $f_{Y,\Delta,\boldsymbol{\alpha}_1}(t,1) = f_{Y,\Delta,\boldsymbol{\alpha}_2}(t,1)$ for all $t$, where $\boldsymbol{\alpha}_j=(\theta_j,\theta_{Tj},\theta_{Cj})^\top$, $j=1,2$.  Then, 
$$ 1 = \lim_{t \rightarrow a} \frac{f_{Y,\Delta,\boldsymbol{\alpha}_1}(t,1)}{f_{Y,\Delta,\boldsymbol{\alpha}_2}(t,1)} = 
\lim_{t \rightarrow a} \frac{f_{T,\theta_{T1}}(t)}{f_{T,\theta_{T2}}(t)}. $$
It follows from condition (C1) that $\theta_{T1}=\theta_{T2}$.  In a similar way we can show that $\theta_{C1}=\theta_{C2}$.   Finally, in order to show that $\theta_1=\theta_2$, note that $F_{Y,\boldsymbol{\alpha}_j}(t) = F_{T,\theta_{T1}}(t) + F_{C,\theta_{C1}}(t) - \C_{\theta_j}(F_{T,\theta_{T1}}(t),F_{C,\theta_{C1}}(t))$ and that $F_{Y,\boldsymbol{\alpha}_1}(t) = F_{Y,\boldsymbol{\alpha}_2}(t)$ for all $t$.  Hence, since the copula is unique, it follows that $\theta_1=\theta_2$.  \hfill $\Box$
\vspace*{.5cm}

\noindent
{\bf Proof of Theorem \ref{condC1}.}  Consider first the log-normal density for $T$ depending on $\boldsymbol{\theta}_T=(\mu,\sigma)$.  Then,
$$ \lim_{t \rightarrow 0} \frac{f_{T,\mu_1,\sigma_1}(t)}{f_{T,\mu_2,\sigma_2}(t)} 
= \lim_{t \rightarrow 0} \frac{\frac{1}{t\sigma_1} \phi\Big(\frac{\log t - \mu_1}{\sigma_1}\Big)}{\frac{1}{t\sigma_2} \phi\Big(\frac{\log t - \mu_2}{\sigma_2}\Big)}
= \lim_{t' \rightarrow -\infty} \frac{\sigma_2}{\sigma_1} \cdot \frac{\phi\Big(\frac{t' - \mu_1}{\sigma_1}\Big)}{\phi\Big(\frac{t' - \mu_2}{\sigma_2}\Big)}, $$
and it is easily seen that this can only be equal to 1 when $\mu_1=\mu_2$ and $\sigma_1=\sigma_2$.  The same is true when taking the limit for $t$ going to $\infty$. 

Similarly, for the log-Student-t, we have
\begin{align}
& \lim_{t \rightarrow 0,\infty} \frac{f_{T,\nu_1,\mu_1,\sigma_1}(t)}{f_{T,\nu_2,\mu_2,\sigma_2}(t)} = \frac{c_{\nu_1,\sigma_1}}{c_{\nu_2,\sigma_2}} \lim_{t \rightarrow 0,\infty} \frac{\big(1+\frac{1}{\nu_1}(\frac{\log t-\mu_1}{\sigma_1})^2 \big)^{-\frac{\nu_1+1}{2}}}{\big(1+\frac{1}{\nu_2}(\frac{\log t-\mu_2}{\sigma_2})^2 \big)^{-\frac{\nu_2+1}{2}}}, \label{studentt}
\end{align}
where 
$c_{\nu,\sigma} = \Gamma(\frac{\nu+1}{2}) / \big[\sigma \sqrt{\nu \pi} \Gamma(\frac{\nu}{2})]$ and $\Gamma$ is the gamma-function.  It is easily seen that the limit in (\ref{studentt}) is equal to 1 if and only if $(\nu_1,\mu_1,\sigma_1)=(\nu_2,\mu_2,\sigma_2)$. 

This is also the case for the Weibull density, since 
\begin{align} \label{Weib}
& \lim_{t \rightarrow 0,\infty} \frac{f_{T,\lambda_1,\rho_1}(t)}{f_{T,\lambda_2,\rho_2}(t)} = \frac{\lambda_1\rho_1}{\lambda_2 \rho_2} \lim_{t \rightarrow 0,\infty} \frac{t^{\rho_1-1} \exp(-\lambda_1 t^{\rho_1})}{t^{\rho_2-1} \exp(-\lambda_2 t^{\rho_2})}, 
\end{align}
and this equals one only if $\rho_1=\rho_2$ and $\lambda_1=\lambda_2$.  

Finally, for the log-logistic density, the limit equals
$$ \lim_{t \rightarrow 0,\infty} \frac{f_{T,\lambda_1,\kappa_1}(t)}{f_{T,\lambda_2,\kappa_2}(t)} = \frac{\kappa_1 \lambda_1^{\kappa_1}}{\kappa_2 \lambda_2^{\kappa_2}} \lim_{t \rightarrow 0,\infty} \frac{t^{\kappa_1-1} [1+(\lambda_2 t)^{\kappa_2}]^2}{t^{\kappa_2-1} [1+(\lambda_1 t)^{\kappa_1}]^2}, $$
and again this can only be equal to one when $\kappa_1=\kappa_2$ and $\lambda_1=\lambda_2$.  Hence, condition (C1) is satisfied for each of these densities.  \hfill $\Box$
\vspace*{.5cm}

\noindent
{\bf Proof of Lemma \ref{lem1}.} We will use the abbreviated notation $u_t = F_{T,\theta_T}(t)$ and $v_t = F_{C,\theta_C}(t)$. 
Note that $\psi(1)=0$, $\lim_{t \rightarrow 0} \psi^{-1}(t) =1$ and $\lim_{u \rightarrow 1} \psi'(u) = c \in (-\infty,0)$.  Hence,
\begin{align*}
\hspace*{2.7cm} \lim_{t \rightarrow \infty} h_{T|C,\theta}(u_t|v_t) &= \lim_{t \rightarrow \infty} \frac{\psi'(v_t)}{\psi'\big(\psi^{-1}(\psi(u_t)+\psi(v_t))\big)}  =  \frac{c}{c}=1. \hspace*{2.1cm} \Box 
\end{align*}


\vspace*{.5cm}

\noindent
{\bf Proof of Theorem \ref{spec}.}  We start with the Frank copula,.  
Straightforward calculations show that $\lim_{u \rightarrow 1} \psi' (u) = \theta e^{-\theta} /(e^{-\theta}-1) < 0$ for $\theta \neq 0$, and hence we know from Lemma \ref{lem1} that $\lim_{t \rightarrow \infty} h_{T|C}(F_T(t)|F_C(t)) = 1$.  Hence, condition (C2a) can only be satisfied if $h_{T|C}(F_T(t)|F_C(t))$ converges to zero for $t \rightarrow 0$.   Some straightforward but tedious calculations show that 
\begin{align*}
& \lim_{t \rightarrow 0} h_{T|C}(F_T(t)|F_C(t)) \\
& = \lim_{t \rightarrow 0} \frac{\psi'(F_C(t))}{\psi'\big(\psi^{-1}\big(\psi(F_T(t)) + \psi(F_C(t))\big)\big)} \\
& =  \lim_{t \rightarrow 0} \frac{e^{-\theta F_C(t)} (e^{-\theta F_T(t)}-1)}{(e^{-\theta F_C(t)}-1)(e^{-\theta F_T(t)}-1) + e^{-\theta}-1} = 0
\end{align*}
for $\theta \neq 0$.  Hence, condition (C2a) is satisfied.  Similarly it can be shown that condition (C2b) is satisfied.    

For the Gumbel family we have that $\lim_{u \rightarrow 1} \psi' (u) = 0$, and hence Lemma \ref{lem1} is not applicable.   Hence, we calculate both the limit for $t$ going to 0 and infinity (see \cite{Aas09} for the formula of $h_{T|C}(F_T(t)|F_C(t))$ for the Gumbel family):
\begin{align*}
& \lim_{t \rightarrow 0,\infty} h_{T|C}(F_T(t)|F_C(t)) \\
& = \lim_{t \rightarrow 0,\infty} \Big\{1+(-\log F_C(t))^{-\theta} (-\log F_T(t))^\theta\Big\}^{-1+1/\theta} \\
& \hspace*{.5cm} \times \lim_{t \rightarrow 0,\infty} \exp \Big\{-\big[(-\log F_T(t))^\theta + (-\log F_C(t))^\theta\big]^{1/\theta} - \log F_C(t)\Big\} .
\end{align*}
\tcb Under the assumption that $\log F_T(t)/\log F_C(t) \rightarrow c$ for some $0<c<\infty$ when $t$ tends to zero, \tcb the exponential factor above tends to 0 when $t$ tends to 0, and tends to 1 when $t$ tends to infinity, whereas the factor in front of this exponential factor converges to some constant in the interval $[0,1]$, depending on the limit of $\log F_T(t)/\log F_C(t)$ (for $t$ tending to 0 or infinity).  This shows that the product of the two limits equals 0 when $t$ tends to 0, whereas when $t$ tends to infinity the limit can be zero but it can also be strictly positive depending on the limit of $\log F_T(t)/\log F_C(t)$ for $t$ growing to infinity. 

Finally, we consider the Gaussian copula.   Note that 
\begin{align*}
& P\big(\Phi^{-1}(F_T(T)) \le t,\Phi^{-1}(F_C(C))\le c \big) \\
& = P\big(T \le F_T^{-1}(\Phi(t)), C \le F_C^{-1}(\Phi(c)) \big) = \Phi_\theta(t,c), 
\end{align*}
and hence $\Phi^{-1}(F_T(T)) | \Phi^{-1}(F_C(C)) \sim N(\theta \Phi^{-1}(F_C(C)), 1-\theta^2)$.  It follows that (we omit the parameters $\theta, \theta_T$ and $\theta_C$ for simplicity)
\begin{align*}
& h_{T|C}(F_T(t)|F_C(t)) = F_{T|C}(t|t) \\
& = P\Big(\Phi^{-1}(F_T(T)) \le \Phi^{-1}(F_T(t)) \, \Big| \, \Phi^{-1}(F_C(C)) = \Phi^{-1}(F_C(t))\Big) \\
& = \Phi\Big(\frac{\Phi^{-1}(F_T(t)) - \theta \Phi^{-1}(F_C(t))}{(1-\theta^2)^{1/2}}\Big) \\
& = \Phi\Big(\frac{A_{\theta,F_T,F_C}(t)}{(1-\theta^2)^{1/2}}\Big).
\end{align*}
Since $A_{\theta,F_T,F_C}(t)$ tends to $-\infty$ either when $t$ tends to 0 or to $\infty$, it follows that $h_{T|C}(F_T(t)|F_C(t))$ tends to 0 either when $t$ tends to 0 or to $\infty$.  Hence, condition $(C2a)$ is satisfied.  In a similar way we can prove condition $(C2b)$. \hfill $\Box$
\vspace*{.5cm}

\noindent
{\bf Proof of Theorem \ref{Clay}.} Suppose that $\lim_{t \rightarrow 0} F_{T,\theta_T}(t)/F_{C,\theta_C}(t) = \infty$ (the case where the limit equals 0 can be handled similarly).  Then, it follows from (\ref{hCT}) that $\lim_{t \rightarrow 0} h_{C|T,\theta}(F_{C,\theta_C}(t)|F_{T,\theta_T}(t))=0$.  Hence, using similar arguments as in the proof of Theorem \ref{ident}, it follows from condition (C1) that $\theta_T$ is identifiable.   From the formula of $f_{Y,\Delta}(\cdot,1)$ given in (\ref{resultsYdelta}) it then follows that the function $t \rightarrow h_{C|T,\theta}(F_{C,\theta_C}(t)|F_{T,\theta_T}(t))$ is identifiable.

Next, note that for $t$ large enough, $F_T(t)^\theta/F_{C,\theta_C}(t)^\theta - F_T(t)^\theta$ is close to zero (we omit $\theta_T$ since it is identifiable), and hence a Taylor expansion can be used to write
\begin{align*}
& \log h_{C|T,\theta}(F_{C,\theta_C}(t)|F_T(t)) \\
& = -\frac{\theta+1}{\theta} \log \Big(1+ \frac{F_T(t)^\theta}{F_{C,\theta_C}(t)^\theta} - F_T(t)^\theta\Big) \\
& = -\frac{\theta+1}{\theta} \sum_{k=1}^\infty \frac{(-1)^{k-1}}{k} \big\{F_T(t)^\theta (F_{C,\theta_C}(t)^{-\theta}-1) \big\}^k
\end{align*}
for $t$ large.   This is a polynomial in $u_t=F_T(t)$.  Hence, for two sets of parameters $(\theta,\theta_C)$ and $(\theta^*,\theta_C^*)$ we have that 
$$ \frac{\theta+1}{\theta} \sum_{k=1}^\infty \frac{(-1)^{k-1}}{k} \big(F_{C,\theta_C}(t)^{-\theta}-1\big)^k u_t^{\theta k} = \frac{\theta^*+1}{\theta^*} \sum_{k=1}^\infty \frac{(-1)^{k-1}}{k} \big(F_{C,\theta_C^*}(t)^{-\theta^*}-1\big)^k u_t^{\theta^* k} $$
for $t$ large.   And this is only possible if $\theta=\theta^*$ and $\theta_C=\theta_C^*$.  \hfill $\Box$

\section*{Supplement} \label{supp}

\setcounter{section}{1}
\renewcommand\theequation{\Alph{section}.\arabic{equation}}

\noindent 
\subsection{Results for Scenario 2}
The results from Scenario 2 are shown in Tables \ref{sim2:frank}, \ref{sim2:clayton},
\ref{sim2:gumbel} and \ref{sim2:gauss} for the Frank, Clayton, Gumbel and Gauss copula, respectively. Compared to Scenario 1 the positive bias for the Frank copula is even higher under Scenario 2 for low dependence. In the other copula cases we also see that there is a slightly higher bias in the low dependence case compared to Scenario 1. However for stronger dependences the performance is satisfactory even for sample sizes as low as 200.  

\begin{table}[ht]
	\centering
	\begin{footnotesize}
		\begin{tabular}{rrrrrrr}
			\hline
			$\tau=.2, n=200$&  $\mu_T$ &  $\log(\sigma_T)$ &  $\mu_C$ &  $\log(\sigma_C)$ &  
			$\mbox{logit}(\tau)$ & $\tau$\\
			
			\hline
			average.estimate & 2.43 & -0.06 & 1.98 & -0.69 & -0.51 & 0.38 \\
			sd.of.average.estimate & 0.01 & 0.01 & 0.00 & 0.01 & 0.06 & 0.01 \\
			average.bias & -0.07 & -0.06 & -0.02 & 0.00 & 0.88 & 0.18 \\
			RMSE & 0.17 & 0.13 & 0.04 & 0.06 & 1.08 & 0.23 \\
			\hline
			$n=500$ &  $\mu_T$ &  $\log(\sigma_T)$ &  $\mu_C$ &  $\log(\sigma_C)$ &  $\mbox{logit}(\tau)$ & $\tau$\\
			\hline
			average.estimate      & 2.44 & -0.04 & 1.99 & -0.69 & -0.59 & 0.36 \\
			sd.of.average.estimate      & 0.01 & 0.01 & 0.00 & 0.00 & 0.05 & 0.01 \\
			average.bias      & -0.06 & -0.04 & -0.01 & -0.00 & 0.80 & 0.16 \\
			RMSE      & 0.13 & 0.10 & 0.03 & 0.04 & 0.95 & 0.19 \\
			\hline
			$n=1000$ &  $\mu_T$ &  $\log(\sigma_T)$ &  $\mu_C$ &  $\log(\sigma_C)$ &  $\mbox{logit}(\tau)$ & $\tau$\\
			\hline
			average.estimate       & 2.18 & -0.01 & 1.99 & -1.38 & -0.73 & 0.33 \\
			sd.of.average.estimate       & 0.00 & 0.00 & 0.00 & 0.00 & 0.02 & 0.00 \\
			average.bias       & -0.01 & -0.01 & -0.01 & 0.01 & 0.66 & 0.13 \\
			RMSE       & 0.04 & 0.04 & 0.01 & 0.03 & 0.68 & 0.13 \\
			\hline
			$\tau=.5, n=200$&  $\mu_T$ &  $\log(\sigma_T)$ &  $\mu_C$ &  $\log(\sigma_C)$ &  
			$\mbox{logit}(\tau)$ & $\tau$\\ 
			\hline
			average.estimate      & 2.52 & -0.00 & 2.00 & -0.69 & -0.02 & 0.49 \\
			sd.of.average.estimate      & 0.02 & 0.01 & 0.00 & 0.01 & 0.07 & 0.01 \\
			average.bias      & 0.02 & -0.00 & -0.00 & 0.00 & -0.02 & -0.01 \\
			RMSE      & 0.22 & 0.14 & 0.04 & 0.06 & 0.66 & 0.15 \\
			\hline
			$n=500$ &  $\mu_T$ &  $\log(\sigma_T)$ &  $\mu_C$ &  $\log(\sigma_C)$ &  $\mbox{logit}(\tau)$ & $\tau$\\
			\hline
			average.estimate      & 2.52 & 0.00 & 2.00 & -0.69 & -0.18 & 0.46 \\
			sd.of.average.estimate      & 0.02 & 0.01 & 0.00 & 0.00 & 0.08 & 0.02 \\
			average.bias      & 0.02 & 0.00 & 0.00 & 0.00 & -0.18 & -0.04 \\
			RMSE      & 0.20 & 0.13 & 0.03 & 0.04 & 0.78 & 0.17 \\
			\hline
			$n=1000$ &  $\mu_T$ &  $\log(\sigma_T)$ &  $\mu_C$ &  $\log(\sigma_C)$ &  $\mbox{logit}(\tau)$ & $\tau$\\
			\hline
			average.estimate      & 2.20 & 0.00 & 2.00 & -1.38 & -0.01 & 0.50 \\
			sd.of.average.estimate      & 0.00 & 0.00 & 0.00 & 0.00 & 0.02 & 0.00 \\
			average.bias      & -0.00 & 0.00 & -0.00 & 0.00 & -0.01 & -0.00 \\
			RMSE      & 0.05 & 0.04 & 0.01 & 0.04 & 0.22 & 0.05 \\
			\hline
			$\tau=.8, n=200$&  $\mu_T$ &  $\log(\sigma_T)$ &  $\mu_C$ &  $\log(\sigma_C)$ &  
			$\mbox{logit}(\tau)$ & $\tau$\\
			\hline
			average.estimate      & 2.52 & -0.00 & 2.00 & -0.69 & 1.54 & 0.81 \\
			sd.of.average.estimate      & 0.02 & 0.01 & 0.00 & 0.01 & 0.06 & 0.01 \\
			average.bias      & 0.02 & -0.00 & 0.00 & -0.00 & 0.15 & 0.01 \\
			RMSE      & 0.21 & 0.14 & 0.03 & 0.07 & 0.60 & 0.07 \\
			\hline
			$n=500$ &  $\mu_T$ &  $\log(\sigma_T)$ &  $\mu_C$ &  $\log(\sigma_C)$ &  $\mbox{logit}(\tau)$ & $\tau$\\
			\hline
			average.estimate      & 2.52 & 0.00 & 2.00 & -0.70 & 1.40 & 0.80 \\
			sd.of.average.estimate      & 0.01 & 0.01 & 0.00 & 0.00 & 0.03 & 0.00 \\
			average.bias      & 0.02 & 0.00 & 0.00 & -0.01 & 0.02 & -0.00 \\
			RMSE      & 0.14 & 0.09 & 0.03 & 0.04 & 0.30 & 0.05 \\
			\hline
			$n=1000$ &  $\mu_T$ &  $\log(\sigma_T)$ &  $\mu_C$ &  $\log(\sigma_C)$ &  $\mbox{logit}(\tau)$ & $\tau$\\
			\hline
			average.estimate      & 2.52 & 0.01 & 2.00 & -0.69 & 1.38 & 0.80 \\
			sd.of.average.estimate      & 0.01 & 0.01 & 0.00 & 0.00 & 0.02 & 0.00 \\
			average.bias      & 0.02 & 0.01 & 0.00 & -0.00 & -0.01 & -0.00 \\
			RMSE      & 0.10 & 0.07 & 0.02 & 0.03 & 0.17 & 0.03 \\[-.1cm]
			\hline
		\end{tabular}
	\end{footnotesize}
	\caption{Simulation results for the case of a Frank copula under Scenario 2.}
	\label{sim2:frank}
\end{table}

\begin{table}[ht]
	\centering
	\begin{footnotesize}
		\begin{tabular}{rrrrrrr}
			\hline
			$\tau=.2, n=200$&  $\mu_T$ &  $\log(\sigma_T)$ &  $\mu_C$ &  $\log(\sigma_C)$ &  
			$\mbox{logit}(\tau)$ & $\tau$\\
			\hline
			average.estimate & 2.45 & -0.04 & 1.98 & -0.67 & -1.60 & 0.27 \\
			sd.of.average.estimate & 0.02 & 0.01 & 0.01 & 0.01 & 0.20 & 0.02 \\
			average.bias & -0.05 & -0.04 & -0.02 & 0.02 & -0.21 & 0.07 \\
			RMSE & 0.20 & 0.14 & 0.07 & 0.07 & 1.97 & 0.22 \\
			\hline
			$n=500$ &  $\mu_T$ &  $\log(\sigma_T)$ &  $\mu_C$ &  $\log(\sigma_C)$ &  $\mbox{logit}(\tau)$ & $\tau$\\
			\hline
			average.estimate  & 2.46 & -0.02 & 1.99 & -0.68 & -1.72 & 0.25 \\
			sd.of.average.estimate  & 0.01 & 0.01 & 0.00 & 0.00 & 0.19 & 0.02 \\
			average.bias  & -0.04 & -0.02 & -0.01 & 0.01 & -0.33 & 0.05 \\
			RMSE  & 0.15 & 0.10 & 0.05 & 0.05 & 1.89 & 0.18 \\
			\hline
			$n=1000$ &  $\mu_T$ &  $\log(\sigma_T)$ &  $\mu_C$ &  $\log(\sigma_C)$ &  $\mbox{logit}(\tau)$ & $\tau$\\
			\hline
			average.estimate  & 2.48 & -0.01 & 2.00 & -0.68 & -1.50 & 0.22 \\
			sd.of.average.estimate  & 0.01 & 0.01 & 0.00 & 0.00 & 0.11 & 0.01 \\
			average.bias  & -0.02 & -0.01 & -0.00 & 0.01 & -0.11 & 0.02 \\
			RMSE  & 0.11 & 0.07 & 0.04 & 0.04 & 1.13 & 0.12 \\
			\hline
			$\tau=.5, n=200$&  $\mu_T$ &  $\log(\sigma_T)$ &  $\mu_C$ &  $\log(\sigma_C)$ &  
			$\mbox{logit}(\tau)$ & $\tau$\\
			\hline
			average.estimate  & 2.52 & 0.01 & 2.00 & -0.69 & -0.10 & 0.49 \\
			sd.of.average.estimate  & 0.02 & 0.01 & 0.00 & 0.01 & 0.08 & 0.01 \\
			average.bias  & 0.02 & 0.01 & 0.00 & -0.00 & -0.10 & -0.01 \\
			RMSE  & 0.20 & 0.14 & 0.05 & 0.07 & 0.85 & 0.15 \\
			\hline
			$n=500$ &  $\mu_T$ &  $\log(\sigma_T)$ &  $\mu_C$ &  $\log(\sigma_C)$ &  $\mbox{logit}(\tau)$ & $\tau$\\
			\hline
			average.estimate  & 2.50 & -0.01 & 2.00 & -0.69 & 0.04 & 0.51 \\
			sd.of.average.estimate  & 0.01 & 0.01 & 0.00 & 0.00 & 0.04 & 0.01 \\
			average.bias  & -0.00 & -0.01 & 0.00 & -0.00 & 0.04 & 0.01 \\
			RMSE  & 0.14 & 0.10 & 0.03 & 0.05 & 0.42 & 0.10 \\
			\hline
			$n=1000$ &  $\mu_T$ &  $\log(\sigma_T)$ &  $\mu_C$ &  $\log(\sigma_C)$ &  $\mbox{logit}(\tau)$ & $\tau$\\
			\hline
			average.estimate  & 2.49 & -0.01 & 2.00 & -0.69 & 0.04 & 0.51 \\
			sd.of.average.estimate  & 0.01 & 0.01 & 0.00 & 0.00 & 0.03 & 0.01 \\
			average.bias  & -0.01 & -0.01 & -0.00 & -0.00 & 0.04 & 0.01 \\
			RMSE  & 0.10 & 0.07 & 0.02 & 0.03 & 0.29 & 0.07 \\
			\hline
			$\tau=.8, n=200$&  $\mu_T$ &  $\log(\sigma_T)$ &  $\mu_C$ &  $\log(\sigma_C)$ &  
			$\mbox{logit}(\tau)$ & $\tau$\\
			\hline
			average.estimate  & 2.50 & -0.01 & 2.00 & -0.69 & 1.52 & 0.81 \\
			sd.of.average.estimate  & 0.02 & 0.01 & 0.00 & 0.01 & 0.05 & 0.01 \\
			average.bias  & -0.00 & -0.01 & -0.00 & -0.00 & 0.13 & 0.01 \\
			RMSE  & 0.20 & 0.15 & 0.04 & 0.06 & 0.51 & 0.07 \\
			\hline
			$n=500$ &  $\mu_T$ &  $\log(\sigma_T)$ &  $\mu_C$ &  $\log(\sigma_C)$ &  $\mbox{logit}(\tau)$ & $\tau$\\
			\hline
			average.estimate  & 2.50 & -0.02 & 2.00 & -0.70 & 1.43 & 0.80 \\
			sd.of.average.estimate  & 0.01 & 0.01 & 0.00 & 0.00 & 0.02 & 0.00 \\
			average.bias  & -0.00 & -0.02 & 0.00 & -0.01 & 0.04 & 0.00 \\
			RMSE  & 0.11 & 0.08 & 0.02 & 0.04 & 0.23 & 0.04 \\
			\hline
			$n=1000$ &  $\mu_T$ &  $\log(\sigma_T)$ &  $\mu_C$ &  $\log(\sigma_C)$ &  $\mbox{logit}(\tau)$ & $\tau$\\
			\hline
			average.estimate & 2.50 & -0.01 & 2.00 & -0.70 & 1.40 & 0.80 \\
			sd.of.average.estimate & 0.01 & 0.01 & 0.00 & 0.00 & 0.02 & 0.00 \\
			average.bias & -0.00 & -0.01 & -0.00 & -0.01 & 0.01 & 0.00 \\
			RMSE & 0.08 & 0.06 & 0.02 & 0.03 & 0.16 & 0.02 \\[-.1cm]
			\hline
		\end{tabular}
	\end{footnotesize}
	\caption{Simulation results for the case of a Clayton copula under Scenario 2.}
	\label{sim2:clayton}
\end{table}

\begin{table}[ht]
	\centering
	\begin{footnotesize}
		\begin{tabular}{lrrrrrr}
			\hline
			$\tau=.2, n=200$&  $\mu_T$ &  $\log(\sigma_T)$ &  $\mu_C$ &  $\log(\sigma_C)$ &  $\mbox{logit}(\tau)$ & $\tau$\\
			\hline
			average.estimate & 2.46 & -0.04 & 1.99 & -0.69 & -1.76 & 0.26 \\
			sd.of.average.estimate & 0.02 & 0.02 & 0.00 & 0.01 & 0.24 & 0.02 \\
			average.bias & -0.04 & -0.04 & -0.01 & 0.00 & -0.37 & 0.06 \\
			RMSE & 0.22 & 0.19 & 0.05 & 0.07 & 2.42 & 0.22 \\
			\hline
			$n=500$ &  $\mu_T$ &  $\log(\sigma_T)$ &  $\mu_C$ &  $\log(\sigma_C)$ &  $\mbox{logit}(\tau)$ & $\tau$\\
			\hline
			average.estimate  & 2.48 & -0.02 & 1.99 & -0.69 & -1.45 & 0.23 \\
			sd.of.average.estimate  & 0.01 & 0.01 & 0.00 & 0.00 & 0.12 & 0.01 \\
			average.bias  & -0.02 & -0.02 & -0.01 & 0.00 & -0.06 & 0.03 \\
			RMSE  & 0.14 & 0.10 & 0.03 & 0.03 & 1.20 & 0.11 \\
			\hline
			$n=1000$ &  $\mu_T$ &  $\log(\sigma_T)$ &  $\mu_C$ &  $\log(\sigma_C)$ &  $\mbox{logit}(\tau)$ & $\tau$\\
			\hline
			average.estimate  & 2.49 & -0.01 & 2.00 & -0.69 & -1.38 & 0.21 \\
			sd.of.average.estimate  & 0.01 & 0.01 & 0.00 & 0.00 & 0.05 & 0.01 \\
			average.bias  & -0.01 & -0.01 & -0.00 & 0.00 & 0.01 & 0.01 \\
			RMSE  & 0.09 & 0.07 & 0.03 & 0.02 & 0.49 & 0.08 \\
			\hline
			\hline
			$ \tau=.5, n=200$ &  $\mu_T$ &  $\log(\sigma_T)$ &  $\mu_C$ &  $\log(\sigma_C)$ &  $\mbox{logit}(\tau)$ & $\tau$\\
			\hline
			average.estimate & 2.50 & -0.03 & 2.00 & -0.68 & 0.15 & 0.53 \\
			sd.of.average.estimate & 0.03 & 0.02 & 0.00 & 0.01 & 0.08 & 0.02 \\
			average.bias & 0.00 & -0.03 & 0.00 & 0.01 & 0.15 & 0.03 \\
			RMSE & 0.26 & 0.20 & 0.05 & 0.06 & 0.86 & 0.18 \\
			\hline
			$n=500$ &  $\mu_T$ &  $\log(\sigma_T)$ &  $\mu_C$ &  $\log(\sigma_C)$ &  $\mbox{logit}(\tau)$ & $\tau$\\
			\hline
			average.estimate  & 2.49 & -0.01 & 2.00 & -0.69 & 0.06 & 0.51 \\
			sd.of.average.estimate  & 0.01 & 0.01 & 0.00 & 0.00 & 0.04 & 0.01 \\
			average.bias  & -0.01 & -0.01 & -0.00 & -0.00 & 0.06 & 0.01 \\
			RMSE  & 0.15 & 0.10 & 0.03 & 0.04 & 0.42 & 0.10 \\
			\hline
			$n=1000$ &  $\mu_T$ &  $\log(\sigma_T)$ &  $\mu_C$ &  $\log(\sigma_C)$ &  $\mbox{logit}(\tau)$ & $\tau$\\
			\hline
			average.estimate  & 2.50 & -0.00 & 2.00 & -0.69 & 0.03 & 0.51 \\
			sd.of.average.estimate  & 0.01 & 0.01 & 0.00 & 0.00 & 0.03 & 0.01 \\
			64
			average.bias  & -0.00 & -0.00 & -0.00 & 0.00 & 0.03 & 0.01 \\
			RMSE  & 0.10 & 0.07 & 0.02 & 0.03 & 0.29 & 0.07 \\
			\hline
			$ \tau=.8, n=200$ &  $\mu_T$ &  $\log(\sigma_T)$ &  $\mu_C$ &  $\log(\sigma_C)$ &  $\mbox{logit}(\tau)$ & $\tau$\\
			\hline
			average.estimate & 2.49 & -0.03 & 2.00 & -0.68 & 1.53 & 0.81 \\
			sd.of.average.estimate & 0.02 & 0.02 & 0.00 & 0.01 & 0.06 & 0.01 \\
			average.bias & -0.01 & -0.03 & 0.00 & 0.01 & 0.14 & 0.01 \\
			RMSE & 0.21 & 0.17 & 0.04 & 0.06 & 0.57 & 0.07 \\
			\hline
			$n=500$ &  $\mu_T$ &  $\log(\sigma_T)$ &  $\mu_C$ &  $\log(\sigma_C)$ &  $\mbox{logit}(\tau)$ & $\tau$\\
			\hline
			average.estimate  & 2.51 & -0.01 & 2.00 & -0.70 & 1.44 & 0.80 \\
			sd.of.average.estimate  & 0.01 & 0.01 & 0.00 & 0.00 & 0.03 & 0.00 \\
			average.bias  & 0.01 & -0.01 & 0.00 & -0.01 & 0.05 & 0.00 \\
			RMSE  & 0.15 & 0.11 & 0.02 & 0.04 & 0.28 & 0.04 \\
			\hline
			$n=1000$ &  $\mu_T$ &  $\log(\sigma_T)$ &  $\mu_C$ &  $\log(\sigma_C)$ &  $\mbox{logit}(\tau)$ & $\tau$\\
			\hline
			average.estimate  & 2.50 & -0.01 & 2.00 & -0.69 & 1.42 & 0.80 \\
			sd.of.average.estimate  & 0.01 & 0.01 & 0.00 & 0.00 & 0.02 & 0.00 \\
			average.bias  & -0.00 & -0.01 & 0.00 & 0.00 & 0.03 & 0.00 \\
			RMSE  & 0.11 & 0.09 & 0.02 & 0.03 & 0.21 & 0.03 \\[-.1cm]
			\hline
		\end{tabular}
	\end{footnotesize}
	\caption{Simulation results for the case of a Gumbel copula under Scenario 2.}
	\label{sim2:gumbel}
\end{table}

\begin{table}[ht]
	\centering
	\begin{footnotesize}
		\begin{tabular}{rrrrrrr}
			\hline
			$\tau=.2, n=200$&  $\mu_T$ &  $\log(\sigma_T)$ &  $\mu_C$ &  $\log(\sigma_C)$ &  $\mbox{logit}(\tau)$ & $\tau$\\
			\hline
			average.estimate & 2.46 & -0.05 & 1.98 & -0.68 & -2.08 & 0.24 \\ 
			sd.of.average.estimate & 0.03 & 0.02 & 0.01 & 0.01 & 0.23 & 0.02 \\ 
			average.bias & -0.04 & -0.05 & -0.02 & 0.02 & -0.69 & 0.04 \\ 
			RMSE & 0.31 & 0.17 & 0.13 & 0.11 & 2.37 & 0.20 \\ 
			\hline
			$n=500$ &  $\mu_T$ &  $\log(\sigma_T)$ &  $\mu_C$ &  $\log(\sigma_C)$ &  $\mbox{logit}(\tau)$ & $\tau$\\
			\hline
			average.estimate & 2.45 & -0.03 & 1.98 & -0.68 & -1.77 & 0.24 \\ 
			sd.of.average.estimate & 0.03 & 0.01 & 0.01 & 0.01 & 0.18 & 0.02 \\ 
			average.bias & -0.05 & -0.03 & -0.02 & 0.01 & -0.38 & 0.04 \\ 
			RMSE & 0.29 & 0.12 & 0.12 & 0.10 & 1.84 & 0.17 \\ 
			\hline
			$n=1000$ &  $\mu_T$ &  $\log(\sigma_T)$ &  $\mu_C$ &  $\log(\sigma_C)$ &  $\mbox{logit}(\tau)$ & $\tau$\\
			\hline
			average.estimate & 2.47 &  -0.01 & 1.98 & -0.68 & -1.69 & 0.21 \\ 
			sd.of.average.estimate & 0.03 & 0.01 & 0.01 & 0.01 & 0.14 & 0.01 \\ 
			average.bias & -0.03 & -0.01 & -0.02 & 0.01 & -0.31 & 0.01 \\ 
			RMSE.2 & 0.26 & 0.09 & 0.11 & 0.10 & 1.39 & 0.11 \\ 
			\hline
			$\tau=.5, n=200$&  $\mu_T$ &  $\log(\sigma_T)$ &  $\mu_C$ &  $\log(\sigma_C)$ &  $\mbox{logit}(\tau)$ & $\tau$\\
			\hline
			average.estimate & 2.49 & -0.01 & 1.99 & -0.68 & -0.08 & 0.50 \\ 
			sd.of.average.estimate & 0.04 & 0.02 & 0.01 & 0.01 & 0.12 & 0.02 \\ 
			average.bias & -0.01 & -0.01 & -0.01 & 0.02 & -0.08 & -0.00 \\ 
			RMSE & 0.34 & 0.17 & 0.12 & 0.12 & 1.21 & 0.18 \\ 
			\hline
			$n=500$ &  $\mu_T$ &  $\log(\sigma_T)$ &  $\mu_C$ &  $\log(\sigma_C)$ &  $\mbox{logit}(\tau)$ & $\tau$\\
			\hline
			average.estimate & 2.48 & -0.01 & 1.99 & -0.68 & -0.05 & 0.50 \\ 
			sd.of.average.estimate & 0.03 & 0.01 & 0.01 & 0.01 & 0.10 & 0.01 \\ 
			average.bias & -0.02 & -0.01 & -0.01 & 0.01 & -0.05 & 0.00 \\ 
			RMSE & 0.29 & 0.12 & 0.11 & 0.10 & 1.00 & 0.13 \\ 
			\hline
			$n=1000$ &  $\mu_T$ &  $\log(\sigma_T)$ &  $\mu_C$ &  $\log(\sigma_C)$ &  $\mbox{logit}(\tau)$ & $\tau$\\
			\hline
			average.estimate & 2.48 & -0.01 & 1.99 & -0.68 & -0.07 & 0.50 \\ 
			sd.of.average.estimate & 0.03 & 0.01 & 0.01 & 0.01 & 0.09 & 0.01 \\ 
			average.bias& -0.02 & -0.01 & -0.01 & 0.02 & -0.07 & -0.00 \\ 
			RMSE & 0.26 & 0.09 & 0.11 & 0.10 & 0.92 & 0.10 \\ 
			\hline
			$\tau=.8, n=200$&  $\mu_T$ &  $\log(\sigma_T)$ &  $\mu_C$ &  $\log(\sigma_C)$ &  $\mbox{logit}(\tau)$ & $\tau$\\
			\hline
			average.estimate & 2.46 & -0.04 & 1.99 & -0.67 & 1.45 & 0.81 \\ 
			sd.of.average.estimate & 0.03 & 0.02 & 0.01 & 0.01 & 0.11 & 0.01 \\ 
			average.bias & 0.04 & -0.04 & -0.01 & 0.02 & 0.06 & 0.01 \\ 
			RMSE.6 & 0.31 & 0.16 & 0.12 & 0.11 & 1.10 & 0.10 \\ 
			\hline
			$n=500$ &  $\mu_T$ &  $\log(\sigma_T)$ &  $\mu_C$ &  $\log(\sigma_C)$ &  $\mbox{logit}(\tau)$ & $\tau$\\
			\hline
			average.estimate & 2.46 & -0.03 & 1.99 & -0.68 & 1.36 & 0.80 \\ 
			sd.of.average.estimate & 0.03 & 0.01 & 0.01 & 0.01 & 0.10 & 0.01 \\ 
			average.bias & -0.04 & -0.03 & -0.01 & 0.01 & -0.02 & 0.00 \\ 
			RMSE & 0.28 & 0.12 & 0.11 & 0.10 & 1.03 & 0.09 \\ 
			\hline
			$n=1000$ &  $\mu_T$ &  $\log(\sigma_T)$ &  $\mu_C$ &  $\log(\sigma_C)$ &  $\mbox{logit}(\tau)$ & $\tau$\\
			\hline
			average.estimate & 2.49 & 0.00 & 1.99 & -0.69 & 1.27 & 0.79 \\ 
			sd.of.average.estimate & 0.03 & 0.01 & 0.01 & 0.01 & 0.10 & 0.01 \\ 
			average.bias & -0.01 & 0.00 & -0.01 & 0.00 & -0.12 & -0.01 \\ 
			RMSE & 0.26 & 0.09 & 0.11 & 0.10 & 1.01 & 0.09 \\ [-.1cm]
			\hline
		\end{tabular}
	\end{footnotesize}
	\caption{Simulation results for the case of a Gauss copula under Scenario 2.}
	\label{sim2:gauss}
\end{table}

\clearpage

\end{document}